\documentclass[12pt]{article}

\usepackage{amsmath,amssymb,amsfonts,amsthm}
\usepackage[T1,T2A]{fontenc}
\usepackage{graphicx}
\usepackage{cite}
\usepackage[all]{xy}
\usepackage[toc,page]{appendix}
\usepackage{hyperref}
\hypersetup{colorlinks=true,  citecolor=red, linkcolor=blue}
\usepackage[mathcal]{eucal}

\allowdisplaybreaks[3]
\newcounter{propositiona}
\newcommand{\propositiona}[1]{\refstepcounter{propositiona}
\noindent
\textbf{Proposition \thepropositiona.}\, {\it #1}}
\newcounter{definitiona}
\newcommand{\definitiona}[1]{\refstepcounter{definitiona}
\noindent
\textbf{Definition \thedefinitiona.}\, #1}
\newcounter{remarka}
\newcommand{\remarka}[1]{\refstepcounter{remarka}
\noindent
\textbf{Remark \theremarka.}\, #1}
\newcounter{examplea}

\newcounter{lemmaa}
\newcommand{\lemmaa}[1]{\refstepcounter{lemmaa}
\noindent
\textbf{Lemma \thelemmaa.}\, {\it #1}}
\newcounter{theorema}
\newcommand{\theorema}[1]{\refstepcounter{theorema}
\noindent
\textbf{Theorem\, \thetheorema.}\, {\it #1}}
\newcounter{corollarya}

\renewcommand{\thefootnote}{\alph{footnote}}

\title{Invariant Reduction for Partial Differential Equations. III: Poisson Brackets}

\author{ \renewcommand{\thefootnote}{\alph{footnote}}
Kostya Druzhkov\hspace{0.1ex}\footnotemark[1] \vspace{0.5cm}\\
\small \emph{Department of Mathematics and Statistics, University of Saskatchewan, Saskatoon, Canada}\vspace{0.2cm}\\
}

\begin{document}

\footnotetext[1]{Electronic mail: konstantin.druzhkov@gmail.com}

\maketitle \numberwithin{equation}{section}
\renewcommand{\thefootnote}{\arabic{footnote}}
%\maketitle \numberwithin{remark}{section}
%\numberwithin{lemma}{section}
%\numberwithin{proposition}{section}

\begin{abstract}
We establish that local Hamiltonian operators admitted by systems of partial differential equations (PDEs) induce Poisson bivectors on suitable finite-dimensional subsystems describing invariant solutions. This correspondence is governed by the mechanism of invariant reduction, which applies uniformly to point, contact, and higher symmetries. We show that the induced Poisson bivectors relate constants of invariant motion to symmetries of such subsystems. Applying the same reduction mechanism to invariant conservation laws, we further deduce that the corresponding Poisson brackets on the reduced systems agree, up to a global sign, with those of the original systems. The results are illustrated by two examples in which the induced Poisson brackets and constants of invariant motion yield integrability of the reduced systems. The construction is independent of the choice of an $\ell$-normal inclusion of a PDE system into jet spaces, with the corresponding odd cotangent covering being $\ell$-normal.
\end{abstract}


\section{Introduction}

Hamiltonian structures play a prominent role in both the theory of infinite-dimensional integrable systems and the geometric theory of partial differential equations (PDEs). Local Hamiltonian operators naturally equip the spaces of conservation laws of the corresponding systems with Poisson brackets. While the intrinsic geometric theory of Poisson brackets on finite-dimensional smooth manifolds is classical, their geometric formulation for differential equations is nontrivial, and existing intrinsic formulations usually rely, unlike those of many other integrability structures, on additional constructions effectively mimicking an $\ell$-normal inclusion of the PDE system into a jet space, provided that such an inclusion exists.
This considerably complicates the treatment of Hamiltonian (or variational Poisson) structures on PDEs, including the natural question of how exactly they behave under maps of differential equations that induce maps between their Cartan distributions.

Among such maps of differential equations, symmetry reductions play a distinguished role. If a system $F = 0$ admits an evolutionary symmetry with characteristic $\varphi$, then the system
$$
F = 0 \hspace{0.1ex},\qquad \varphi = 0\hspace{0.1ex},
$$
describing the corresponding invariant solutions, inherits invariant cohomological structures of the original system via the mechanism of invariant reduction~\cite{DrCh2}. Through this broadly applicable mechanism, such symmetry reductions inherit invariant conservation laws, presymplectic structures, internal Lagrangians~\cite{Druzhkov1}, and other geometric structures that admit intrinsic formulations.

In this paper, we apply the same reduction mechanism to invariant Hamiltonian structures by interpreting them as conservation laws (and symmetries) of cotangent coverings with odd fibers~\cite{KraVer}. Although our approach is also applicable to infinite-dimensional reductions, our main focus is on finite-dimensional
ones, for which we define cotangent equations intrinsically. This allows us to identify the reduced conservation laws as objects on the corresponding reductions of the original system.

\vspace{1ex}

\noindent
\textbf{The main idea.} A local Hamiltonian operator $\nabla$ of a system of differential equations $\mathcal{E}$ gives rise to a conservation law $\mathcal{H}_{\nabla}$ of its cotangent covering $\mathcal{E}^*$ with odd fibers. A local symmetry $X$ (or a commutative algebra of local symmetries) of $\mathcal{E}$ lifts to the cotangent equation $\mathcal{E}^*$, which is endowed with a canonical variational $1$-form. Hence, the system $\mathcal{E}_X$ describing $X$-invariant solutions of $\mathcal{E}$
is covered by the corresponding reduction of the cotangent equation, which inherits the canonical variational $1$-form through the invariant reduction mechanism.
Thus, if $\mathcal{E}_X$ is finite-dimensional, two objects arise that serve as cotangent equations for it. 
Under certain regularity conditions, and taking into account the reduced canonical variational $1$-form, these two objects can be uniquely identified.
If $\nabla$ is $X$-invariant, this identification allows one to interpret the reduction of $\mathcal{H}_{\nabla}$ as a bivector on $\mathcal{E}_X$.
The bivector is Poisson, relating constants of $X$-invariant motion to symmetries of $\mathcal{E}_X$. In terms of the reduction of $X$-invariant conservation laws of $\mathcal{E}$, the resulting Poisson bracket on $\mathcal{E}_X$ agrees (up to sign) with the one defined by $\nabla$ on $\mathcal{E}$.

\vspace{1ex}

In this regard, evolution systems do not play a distinguished role; however, an explicit formula can be written for them in the $(1+1)$-dimensional case, as given in Theorem~\ref{MainTheor}. In that setting, the reductions of conservation laws can be computed using an algorithm provided as \verb|Maple| code in~\cite{DrCh1}. This also makes invariant reduction a useful tool for systematically studying properties of invariant solutions.

There is a substantial literature on related questions, including specific results on finite-dimensional symmetry reductions and the inheritance of Hamiltonian structures for evolution systems (see, e.g.,~\cite{BogoNov, NoMaPiZa, AntFoWo, AntoRau, Tondo, Fordy, FaMaPeZu, FoHu, FoHu1, BlaMaSza}). Our approach is distinguished by its generality and systematic nature: assuming suitable regularity conditions, it applies to $\ell$-normal systems of partial differential equations having $\ell$-normal odd cotangent coverings, with $(1+1)$-dimensional evolution systems included as a particular case. It is also global and geometric. In particular, it does not rely on local coordinates, auxiliary variables for reduced systems, or the existence of Lax pairs. Furthermore, it is canonical in the sense that it is independent of the choice of an $\ell$-normal inclusion of a PDE system into jet spaces (with the corresponding odd cotangent covering being $\ell$-normal). While some aspects of symmetry reductions and Poisson bracket inheritance have been studied previously, the generality, systematic nature, and formulation of our approach -- including Theorem~\ref{MainTheor} -- appear to be novel.

Our mechanism of invariant reduction is related to methods introduced in~\cite{AndFel} and~\cite{AncoGa}. As formulated in those works, these methods are not directly applicable to higher symmetries (and have not been applied to Hamiltonian structures), but they allow one to consider multi-reduction under suitable, in some cases non-solvable, multidimensional Lie algebras of point symmetries. Moreover, following the ideas of this paper, a slight extension of the method in~\cite{AndFel} to the reduction of the invariant part of the zero page of Vinogradov’s $\mathcal{C}$-spectral sequence for a given equation may allow Hamiltonian structure reductions for a wider class of multidimensional Lie algebras of point or contact symmetries.

It is worth separately noting the approach introduced in~\cite{Mokh1, Mokh2} (see also~\cite{Ugaglia}). It deals with Poisson bivectors on subsystems of evolution PDEs that describe stationary points of conservation laws (within suitable classes of initial conditions). Although it does not directly involve Hamiltonian operators, it apparently allows one to obtain the same Poisson bivectors in some situations.

The paper is organized as follows. In Section~\ref{Basnot} we introduce notation and provide basic facts from the geometry of PDEs. Section~\ref{Cotlnorm} recalls the concept of the cotangent covering for an $\ell$-normal system. Here we follow~\cite{KraVer}, as well as in Section~\ref{HamopSect}, where Hamiltonian operators are interpreted as conservation laws (and odd symmetries) of the cotangent equations, and some useful formulas are given. Section~\ref{invredmechsec} recalls the invariant reduction mechanism. In Section~\ref{finitedimcovsec} we intrinsically define the cotangent covering of a finite-dimensional system of differential equations. Section~\ref{finitedimredsec} is the main section. It takes the formulas from Section~\ref{HamopSect}, applies the invariant reduction to them, and interprets the results in terms of Section~\ref{finitedimcovsec}. Section~\ref{Sectexamp} presents illustrative examples.

\textit{All functions and manifolds considered in this paper are assumed to be smooth of class $C^{\infty}$.}

\section{Basic notation and concepts \label{Basnot}}

Let us introduce some notation and briefly recall basic facts from the geometry of differential equations. More details can be found in~\cite{VinKr, KraVer}.

\subsection{Jets}

Let $\pi\colon E^{n+m}\to M^n$ be a smooth real vector
bundle over a smooth manifold $M$. The bundle $\pi$ gives rise to the corresponding jet bundles $\pi_k\colon J^k(\pi)\to M$, the tower
\begin{align*}
\xymatrix{
\ldots \ar[r] & J^3(\pi) \ar[r]^-{\pi_{3, 2}} & J^2(\pi) \ar[r]^-{\pi_{2, 1}} & J^1(\pi) \ar[r]^-{\pi_{1, 0}} & J^0(\pi) = E \ar[r]^-\pi & M,
}
\end{align*}
and its inverse limit $J^{\infty}(\pi)$, which arises with the natural projections 
$\pi_{\infty}\colon J^{\infty}(\pi)\to M$ and $\pi_{\infty,\, k}\colon J^{\infty}(\pi)\to J^k(\pi)$. Denote by $\mathcal{F}(\pi)$ and $\Lambda^*(\pi)$ the algebras of smooth functions and differential forms on $J^{\infty}(\pi)$, respectively.

\vspace{0.5ex}
\noindent
\textbf{Local coordinates.} Suppose $U\subset M$ is a coordinate neighborhood such that the bundle $\pi$
becomes trivial over $U$. Choose local coordinates $x^1$, \ldots, $x^n$ in $U$ and $u^1$, \ldots, $u^m$  
along the fibers of $\pi$ over $U$. It is convenient to introduce a
multi-index $\alpha$ as a formal sum of the form $\alpha = \alpha_1 x^1 + \ldots + \alpha_n x^n = \alpha_i x^i$, where all $\alpha_i$ are non-negative integers; $|\alpha| = \alpha_1 + \ldots + \alpha_n$.
We denote by $u^i_{\alpha}$ the corresponding adapted local coordinates on $J^{\infty}(\pi)$ and call them derivatives. Here $u^i_0 = u^i$. The order of $u^i_{\alpha}$ is $|\alpha|$.

\vspace{0.5ex}
\noindent
\textbf{Cartan distribution.} The main structure on jet manifolds is the Cartan distribution.
Using adapted local coordinates on $J^{\infty}(\pi)$, one can introduce the total derivatives
$$
D_{x^i} = \partial_{x^i} + u^k_{\alpha + x^i}\partial_{u^k_{\alpha}}\qquad\quad i = 1, \ldots, n.
$$
The planes of the Cartan distribution $\mathcal{C}$ on $J^{\infty}(\pi)$ are spanned by the total derivatives. The Cartan distribution is involutive and defines a flat connection on $\pi_{\infty}\colon J^{\infty}(\pi)\to M$.

\vspace{0.5ex}
\noindent
\textbf{Cartan forms.} The Cartan distribution $\mathcal{C}$ determines the ideal $\mathcal{C}\Lambda^*(\pi) \subset \Lambda^*(\pi)$
generated by Cartan (or contact) forms, i.e., differential forms that vanish on all planes of the Cartan distribution.
A Cartan $1$-form $\omega\in\mathcal{C}\Lambda^1(\pi)$ can be written as a finite sum
$$
\omega = \omega_i^{\alpha}\theta^i_{\alpha}\,,\qquad\ \theta^i_{\alpha} = du^i_{\alpha} - u^i_{\alpha + x^k}dx^k
$$
in adapted local coordinates. The coefficients $\omega_i^{\alpha}$ are smooth functions of adapted coordinates.

\vspace{0.5ex}
\noindent
\textbf{Infinitesimal symmetries.}
Smooth sections of the pullback bundles $\pi^*_{k}(\pi)\colon \pi^*_{k}(E)\to J^k(\pi)$ determine sections of the pullback $\pi_{\infty}^*(\pi)\colon \pi^*_{\infty}(E)\to J^{\infty}(\pi)$ by means of the projections $\pi_{\infty,\hspace{0.1ex} k}$. We denote by $\varkappa(\pi)$ the $\mathcal{F}(\pi)$-module of such sections of $\pi_{\infty}^*(\pi)$. Each $\chi\in \varkappa(\pi)$ gives rise to the corresponding evolutionary vector field
$$
E_{\chi} = D_{\alpha}(\chi^i)\partial_{u^i_{\alpha}}\,,
$$
where $\chi = (\chi^1$, \ldots, $\chi^m)$ in adapted coordinates, $D_{\alpha} = D_{x^1}^{\ \alpha_1}\circ\ldots\circ D_{x^n}^{\ \alpha_n}$.
Evolutionary vector fields are infinitesimal symmetries of $J^{\infty}(\pi)$. In particular, $\mathcal{L}_{E_{\chi}}\,\mathcal{C}\Lambda^*(\pi)\subset \mathcal{C}\Lambda^*(\pi)$. Here $\mathcal{L}_{E_{\chi}}$ is the corresponding Lie derivative. Elements of $\varkappa(\pi)$ are characteristics of symmetries of $J^{\infty}(\pi)$.

\vspace{0.5ex}
\noindent
\textbf{Horizontal forms.}
Cartan forms allow one to consider the module of horizontal $k$-forms
$$
\Lambda^k_h(\pi) = \Lambda^k(\pi)/\mathcal{C}\Lambda^k(\pi)\,.
$$
The de Rham differential $d$ induces the horizontal differential $d_h\colon \Lambda^k_h(\pi)\to \Lambda^{k+1}_h(\pi)$.
The infinite jet bundle $\pi_{\infty}\colon J^{\infty}(\pi) \to M$ admits the decomposition
$$
\Lambda^1(\pi) = \mathcal{C}\Lambda^1(\pi) \oplus \mathcal{F}(\pi)\!\cdot\!\pi^*_{\infty}(\Lambda^1(M))\,.
$$
\textit{We identify the module of horizontal $k$-forms $\Lambda^k_h(\pi)$ with $\mathcal{F}(\pi)\cdot \pi^*_{\infty}(\Lambda^k(M))$}.
In adapted local coordinates, elements of $\mathcal{F}(\pi)\cdot \pi^*_{\infty}(\Lambda^k(M))$ are generated by the differentials $dx^1, \ldots, dx^n$, while $d_h = dx^i\wedge D_{x^i}$. For example,
$$
d_h (\xi_j dx^j) = dx^i\wedge D_{x^i}(\xi_j) dx^j = D_{x^i}(\xi_j)dx^i\wedge dx^j.
$$

\noindent
\textbf{Euler operator.} Let $\widehat{\varkappa}(\pi)$ be the adjoint module
\begin{align*}
\widehat{\varkappa}(\pi) = \mathrm{Hom}_{\mathcal{F}(\pi)}(\varkappa(\pi), \Lambda^n_h(\pi))\,.
\end{align*}
Denote by $\mathrm{E}$ the Euler operator (variational derivative), $\mathrm{E}\colon \Lambda^n_h(\pi)\to \widehat{\varkappa}(\pi)$.
In adapted coordinates, for $L = \lambda \, dx^1\wedge\ldots\wedge dx^n$ and $\chi\in\varkappa(\pi)$, one has
\begin{align*}
&\mathrm{E}(L)\colon\chi \mapsto \langle \chi, \mathrm{E}(L) \rangle = \chi^i \dfrac{\delta \lambda}{\delta u^i}\, dx^1\wedge\ldots\wedge dx^n\,, \qquad \dfrac{\delta \lambda}{\delta u^i} = \sum_{\alpha} (-1)^{|\alpha|}D_{\alpha}\Big(\dfrac{\partial \lambda}{\partial u^i_{\alpha}}\Big)\,.
\end{align*}
Here $\langle \cdot, \cdot \rangle$ denotes the natural pairing between a module and its adjoint. Note that $\mathrm{E}(L)$ can be identified with the differential form that takes the following form in adapted coordinates:
$$
\mathrm{E}(L) = \theta^i \dfrac{\delta \lambda}{\delta u^i}\wedge dx^1\wedge\ldots\wedge dx^n\,.
$$

\subsection{Differential equations \label{SectionDiffEq}}

Let $\zeta \colon E_1\to M$ be a smooth real vector bundle over the same base as $\pi$. Smooth sections of the pullbacks $\pi^*_{k}(\zeta)$ determine a module of sections of the pullback $\pi^*_{\infty}(\zeta)\colon \pi^*_{\infty}(E_1)\to J^{\infty}(\pi)$. We denote it by $P(\pi)$. Any $F\in P(\pi)$ can be considered a (generally, nonlinear) differential operator $\Gamma(\pi)\to \Gamma(\zeta)$. Then $F = 0$ is a differential equation.
By its \emph{infinite prolongation} we mean the subset $\mathcal{E}\subset J^{\infty}(\pi)$ defined by the infinite system of equations
\begin{align*}
\mathcal{E}\colon\qquad D_{\alpha}(F^i) = 0\,,\qquad |\alpha| \geqslant 0\,.
\end{align*}
Here $F^i$ are components of $F$ in adapted coordinates. We denote $\pi_{\mathcal{E}} = \pi_{\infty}|_{\mathcal{E}}$ and assume that $\pi_{\mathcal{E}}\colon \mathcal{E}\to M$ is surjective. The algebras of smooth functions and differential forms on $\mathcal{E}$ are~given~by
$$
\mathcal{F}(\mathcal{E}) = \mathcal{F}(\pi)/I_{\mathcal{E}}\,,\qquad \Lambda^*(\mathcal{E}) = \bigoplus_{i\geqslant 0} \Lambda^i(\pi)/(I_{\mathcal{E}}\cdot \Lambda^i(\pi) + \Lambda^{i-1}(\pi)\wedge dI_{\mathcal{E}})\,,
$$
respectively.
Here $I_{\mathcal{E}}$ denotes the ideal of the system $\mathcal{E}\subset J^{\infty}(\pi)$, $I_{\mathcal{E}} = \{f\in \mathcal{F}(\pi)\, \colon\, f|_{\mathcal{E}} = 0\}$. Tangent vectors and vector fields on $\mathcal{E}$ are defined in terms of derivations of the algebra $\mathcal{F}(\mathcal{E})$.

\vspace{0.5ex}
\noindent
\textbf{Cartan forms.} The ideal $\mathcal{C}\Lambda^*(\mathcal{E})\subset \Lambda^*(\mathcal{E})$ of Cartan forms on $\mathcal{E}$ is given by $\mathcal{C}\Lambda^*(\pi)|_{\mathcal{E}}$.
It is generated by differential forms that vanish on all planes of the Cartan distribution of $\mathcal{E}$.

\vspace{0.5ex}
\noindent
\textbf{Regularity assumptions.} We say that $\mathcal{E}\subset J^{\infty}(\pi)$ is \emph{regular} if the following conditions hold. 
\begin{enumerate}
\item A function $f\in \mathcal{F}(\pi)$ vanishes on $\mathcal{E}$ if and only if there is a differential operator $\Delta\colon P(\pi)\to \mathcal{F}(\pi)$ of the form $\Delta_i^{\alpha} D_{\alpha}$ ($\mathcal{C}$-differential operator or total differential operator) such that $f = \Delta(F)$. Here, for some integer $k$, the components $\Delta_i^{\alpha}$ with $|\alpha|\leqslant k$ may depend on the independent variables $x^i$, the dependent variables $u^i$, and derivatives up to order $k$, while $\Delta_i^{\alpha} = 0$ for $|\alpha| > k$.
\item $\mathcal{E}$ is representable as the inverse limit\footnote{By this we mean that there exist dg-algebra morphisms $\Lambda^*(\mathcal{E}_k)\to \Lambda^*(\mathcal{E})$ exhibiting $\Lambda^*(\mathcal{E})$ as the corresponding direct limit in the category of dg-algebras.} of a tower of finite-dimensional smooth manifolds
\begin{align}
\xymatrix{
\ldots \ar[r] & \mathcal{E}_{k+1} \ar[r] & \mathcal{E}_{k} \ar[r] & \ldots \ar[r] & \mathcal{E}_0,
}
\label{ETower}
\end{align}
where all maps are surjective submersions.
\end{enumerate}

We consider only regular systems. For simplicity, we assume that the de Rham cohomology groups $H^i_{dR}$ of all systems under consideration are trivial for $i > 0$.

\vspace{0.5ex}
\noindent
\textbf{Infinitesimal symmetries.} A \emph{symmetry} of $\mathcal{E}$ is a vector field $X$ on $\mathcal{E}$ that preserves the Cartan distribution: 
$[X, \mathcal{C}D(\mathcal{E})]\subset \mathcal{C}D(\mathcal{E})$, where $\mathcal{C}D(\mathcal{E})$ denotes the module of \emph{Cartan derivations}, i.e., vector fields on $\mathcal{E}$ whose vectors lie in the respective planes of the Cartan distribution. Two symmetries are equivalent if they differ by a Cartan derivation.

If $\varphi\in \varkappa(\pi)$ is a characteristic such that $E_{\varphi}$ is tangent to $\mathcal{E}$ (i.e., $E_{\varphi}(F)|_{\mathcal{E}} = 0$, or equivalently, $E_{\varphi}(I_{\mathcal{E}})\subset I_{\mathcal{E}}$), then the restriction $E_{\varphi}|_{\mathcal{E}}\colon \mathcal{F}(\mathcal{E})\to \mathcal{F}(\mathcal{E})$ is a (evolutionary) symmetry of $\mathcal{E}$. In this case, the vector field $E_{\varphi}$ can also be called a symmetry of $\mathcal{E}\subset J^{\infty}(\pi)$.
Its characteristic on $\mathcal{E}$, given by $\varphi|_{\mathcal{E}}$, is an element of the kernel of the linearization operator $l_{\mathcal{E}} = l_F|_{\mathcal{E}}\colon \varkappa(\mathcal{E})\to P(\mathcal{E})$, where $\varkappa(\mathcal{E}) = \varkappa(\pi)/I_{\mathcal{E}}\cdot \varkappa(\pi)$, $P(\mathcal{E}) = P(\pi)/I_{\mathcal{E}}\cdot P(\pi)$, and $l_F\colon \varkappa(\pi)\to P(\pi)$ is defined by $l_F(\chi) = E_{\chi}(F)$.

We say that a (regular) system $\mathcal{E}\subset J^{\infty}(\pi)$ is \emph{$\ell$-normal} if every $\mathcal{C}$-differential operator $\Delta\colon P(\mathcal{E})\to \mathcal{F}(\mathcal{E})$ such that $\Delta \circ l_{\mathcal{E}} = 0$ necessarily vanishes.
For instance, all systems of evolution equations are $\ell$-normal.

\vspace{0.5ex}
\noindent
\textbf{$\mathcal{C}$-spectral sequence.} For $p \geqslant 1$, the ideals $\mathcal{C}^p\Lambda^*(\mathcal{E}) = \mathcal{C}^p\Lambda^*(\pi)|_{\mathcal{E}}$ are stable with respect to the de Rham differential, $d(\mathcal{C}^p\Lambda^*(\mathcal{E})) \subset \mathcal{C}^p\Lambda^*(\mathcal{E})$.
The filtration 
$$
\Lambda^{\bullet}(\mathcal{E})\supset \mathcal{C}\Lambda^{\bullet}(\mathcal{E})\supset \mathcal{C}^2\Lambda^{\bullet}(\mathcal{E})\supset \mathcal{C}^3\Lambda^{\bullet}(\mathcal{E})\supset \ldots
$$
of the de Rham complex $\Lambda^{\bullet}(\mathcal{E})$ gives rise to Vinogradov's $\mathcal{C}$-spectral sequence $(E^{p,\hspace{0.1ex} q}_r(\mathcal{E}), d^{p,\hspace{0.1ex} q}_r)$~\cite{Vin, VinKr}.
Here $\mathcal{C}^{k+1}\Lambda^k(\mathcal{E}) = 0$, $E^{p, \hspace{0.1ex} q}_0(\mathcal{E}) = \mathcal{C}^p\Lambda^{p+q}(\mathcal{E})/\mathcal{C}^{p+1}\Lambda^{p+q}(\mathcal{E})$. All differentials $d_r^{p,\hspace{0.1ex} q}$ are induced by the de Rham differential $d$,
$$
d_r^{p,\hspace{0.1ex} q}\colon E^{p, \hspace{0.1ex} q}_r(\mathcal{E}) \to E^{p+r, \hspace{0.1ex} q+1-r}_r(\mathcal{E})\,,\qquad E^{p, \hspace{0.1ex} q}_{r+1}(\mathcal{E}) = \ker d_r^{p,\hspace{0.1ex} q}/ \mathrm{im}\, d_r^{p-r,\hspace{0.1ex} q+r-1}\qquad \text{for}\quad r\geqslant 0\,.
$$
We also use the notation $d_r$ where it does not lead to confusion.

Each element of $E^{p, \hspace{0.1ex} q}_0(\mathcal{E})$ has a unique representative in the restriction of $\mathcal{C}^p\Lambda^p(\pi)\wedge \pi_{\infty}^*(\Lambda^q(M))$.
\textit{We identify elements of $E^{p, \hspace{0.1ex} q}_0(\mathcal{E})$ with their representatives of this form}. A \emph{variational $k$-form} of $\mathcal{E}$ is an element of the group $E^{k,\hspace{0.1ex} n-1}_1(\mathcal{E})$.

\vspace{0.5ex}

\noindent
\textbf{Conservation laws.} A \emph{conservation law} of $\mathcal{E}$ is a variational $0$-form, i.e., an element of the group $E^{0,\hspace{0.1ex} n-1}_1(\mathcal{E})$. Let us denote by $\widehat{P}(\pi)$ the adjoint module
\begin{align*}
\widehat{P}(\pi) = \mathrm{Hom}_{\mathcal{F}(\pi)}(P(\pi), \Lambda^n_h(\pi))\,.
\end{align*}
A \emph{characteristic} of a conservation law~\cite{Olver} is any homomorphism $\psi\in \widehat{P}(\pi)$ such that
\begin{align*}
\langle \psi, F \rangle = d_h\hspace{0.15ex} \mu\,,\qquad \mu\in\Lambda_h^{n-1}(\pi)\,,
\end{align*}
where $\mu|_{\mathcal{E}}\in\Lambda_h^{n-1}(\mathcal{E}) = E_0^{0,\hspace{0.1ex} n-1}(\mathcal{E})$ represents the conservation law.

\vspace{0.5ex}

\noindent
\textbf{Cosymmetries.} Elements of the kernel of the adjoint linearization operator $l_{\mathcal{E}}^{\hspace{0.15ex} *}\colon \widehat{P}(\mathcal{E})\to \widehat{\varkappa}(\mathcal{E})$ are \emph{cosymmetries} of $\mathcal{E}$. Here $l_{\mathcal{E}}^{\hspace{0.15ex} *} = l_{F}^{\hspace{0.15ex} *}|_{\mathcal{E}}$, $\widehat{P}(\mathcal{E}) = \widehat{P}(\pi)/I_{\mathcal{E}}\cdot \widehat{P}(\pi)$, $\widehat{\varkappa}(\mathcal{E}) = \widehat{\varkappa}(\pi)/I_{\mathcal{E}}\cdot \widehat{\varkappa}(\pi)$. The restriction $\overline{\psi} = \psi|_{\mathcal{E}}$ of a characteristic $\psi$ of a conservation law is a cosymmetry. 

More exactly, cosymmetries encode variational $1$-forms. According to the Green formula (the definition of $l_F^{\hspace{0.15ex} *}$), for a cosymmetry $\overline{\psi}\in \ker l_{\mathcal{E}}^{\hspace{0.15ex} *}$, there exists $\omega\in \mathcal{C}\Lambda^1(\mathcal{\pi})\wedge \pi^*_{\infty}(\Lambda^{n-1}(M))$ such that
\begin{align}
\langle l_F(\chi), \psi \rangle = \langle \chi, l_F^{\hspace{0.15ex} *}(\psi) \rangle + d_h (E_{\chi} \lrcorner\, \omega) \qquad \text{for any}\quad \chi\in \varkappa(\pi)\,.
\label{Cosymtoonef}
\end{align}
The differential form $\omega|_{\mathcal{E}}$ represents a variational $1$-form of $\mathcal{E}$. This map from cosymmetries to variational $1$-forms is well-defined and surjective. The correspondence is an isomorphism if $\mathcal{E}$ is $\ell$-normal~\cite{VinKr}.
In this case, $\ker d_1^{\hspace{0.1ex} 0, \hspace{0.1ex} n-1} = 0$ and each conservation law $\xi \in E^{0,\hspace{0.1ex} n-1}_1(\mathcal{E})$ is associated with a unique cosymmetry -- namely, the one corresponding to $d_1\xi$. If $F = 0$ is an Euler--Lagrange equation, then $l_F = l_F^{\hspace{0.15ex} *}$ and cosymmetries of $\mathcal{E}$ can be regarded as its symmetries.

\vspace{0.5ex}
\noindent
\textbf{Presymplectic structures.} A \emph{presymplectic structure} of $\mathcal{E}$ is a $d_1$-closed variational $2$-form, i.e., an element of the kernel of the differential
$$
d_1^{2,\hspace{0.1ex} n-1}\colon E^{2,\hspace{0.1ex} n-1}_1(\mathcal{E})\to E^{3,\hspace{0.1ex} n-1}_1(\mathcal{E}).
$$

\vspace{0.5ex}
\noindent
\textbf{Internal Lagrangian formalism (see~\cite{Druzhkov1}).} 
Suppose that the variational derivative $\mathrm{E}(L)\in \mathcal{C}\Lambda^{n+1}(\pi)$ of a 
horizontal $n$-form $L\in\mathcal{F}(\pi)\cdot \pi_{\infty}^*(\Lambda^n(M))$ vanishes on $\mathcal{E}$.
There exists a form $\omega_L\in\mathcal{C}\Lambda^{n}(\pi)$ such that $d(L + \omega_L) - \mathrm{E}(L) \in\mathcal{C}^2\Lambda^{n+1}(\pi)$. Although such a form $\omega_L$ is defined ambiguously, all differential forms of the form $(L + \omega_L)|_{\mathcal{E}}$ represent the same element of the group
\begin{align}
\dfrac{\{l\in \Lambda^n(\mathcal{E})\, \colon \ dl \in \mathcal{C}^2\Lambda^{n+1}(\mathcal{E})\}}{\mathcal{C}^2\Lambda^{n}(\mathcal{E}) + d(\mathcal{C}\Lambda^{n-1}(\mathcal{E}))}\,.
\label{IntSepLag}
\end{align}
Thus, $L$ defines a unique element of~\eqref{IntSepLag}. Here one can take any $\omega_L \in \mathcal{C}\Lambda^1(\pi)\wedge \pi_{\infty}^*(\Lambda^{n-1}(M))$ such that for every $\chi\in \varkappa(\pi)$, 
$$
\mathcal{L}_{E_{\chi}} L = \langle \chi, \mathrm{E}(L) \rangle + d_h (E_{\chi}\, \lrcorner \, \omega_L)\,.
$$
It can be derived using integration by parts. If $L|_{\mathcal{E}} = 0$, then the corresponding element of~\eqref{IntSepLag} is a variational $1$-form of $\mathcal{E}$.

If $F = 0$ is an Euler--Lagrange equation, then the relation between symmetries and cosymmetries of $\mathcal{E}$ can be described in terms of variational $1$-forms and the presymplectic structure $\Omega$ represented by $d(L + \omega_L)|_{\mathcal{E}}$. The variational $1$-form of $\mathcal{E}$ that corresponds to the cosymmetry arising from a symmetry $X = E_{\varphi}|_{\mathcal{E}}$ is given by the contraction $X \lrcorner\, \Omega$\,.

\section{\label{Cotlnorm} Cotangent coverings of $\ell$-normal systems}

Let us recall the cotangent covering~\cite{KraVer} of an $\ell$-normal system. We assume that $\mathrm{rank}\, \pi = \mathrm{rank}\, \zeta$.

Let $\eta$ be the densitized dual to the bundle $\zeta$, so that the module of its sections reads
\begin{align*}
\Gamma(\eta) = \mathrm{Hom}_{\hspace{0.15ex} C^{\infty}(M)}(\Gamma(\zeta), \Lambda^n(M))\,.
\end{align*}
Denote by $ p = (p_1, \ldots, p_m)$ coordinates along the fibers of $\eta$ over a coordinate neighborhood $U\subset M$ such that both $\pi$ and $\eta$ become trivial over it.
The Whitney sum $\hat{\pi} = \pi \oplus \eta$ gives rise to the corresponding infinite jet bundle\footnote{If $\zeta = \pi$, then $\hat{\pi}_{\infty}$ aligns with the construction of the cotangent bundle to $\pi$ proposed in~\cite{Kuper}.} $\hat{\pi}_{\infty}\colon J^{\infty}(\hat{\pi})\to M$ and the projection
\begin{align}
\hat{\pi}_{\pi}\colon J^{\infty}(\hat{\pi})\to J^{\infty}(\pi)
\label{proj}
\end{align}
with (adapted) coordinates $p_{i\,\alpha}$ along its fibers. In order to describe the invariant reduction of Hamiltonian structures, we
treat $p_{i\,\alpha}$ as odd variables of degree~$1$ and assume that the algebra $\Lambda^*(\hat{\pi})$ is bigraded. For details see Appendix~\ref{App:A}.

Using~\eqref{proj}, we can interpret sections of bundles over $J^{\infty}(\pi)$ as sections of the respective pullback bundles over $J^{\infty}(\hat{\pi})$. In particular, this applies to elements of $\varkappa(\pi)$, $P(\pi)$, and $\widehat{P}(\pi)$. 
Note that the $\mathcal{F}(\hat{\pi})$-module of sections of the pullback $\hat{\pi}_{\infty}^{\hspace{0.1ex} *}(\pi)$ is a direct summand
in $\varkappa(\hat{\pi})$. For $\chi\in \varkappa(\pi)$, the evolutionary vector field $E_{(\chi,\,0)}$
on $J^{\infty}(\hat{\pi})$ has the same form in adapted local coordinates as $E_{\chi}$ on $J^{\infty}(\pi)$.
The lift of the total derivative $D_{x^k}$ to $J^{\infty}(\hat{\pi})$ is the total derivative
\begin{align*}
\mathcal{D}_{x^k} = \partial_{x^k} + u^i_{\alpha + x^k}\partial_{u^i_{\alpha}} + p_{i\,\alpha + x^k}\partial_{p_{i\,\alpha}}\,.
\end{align*}
Then each $\mathcal{C}$-differential operator on $J^{\infty}(\pi)$ can be interpreted as a $\mathcal{C}$-differential operator on $J^{\infty}(\hat{\pi})$. 
We use the following notation on $J^{\infty}(\hat{\pi})$ over $U\subset M$
\begin{align*}
\theta^i_{\alpha} = du^i_{\alpha} - u^i_{\alpha + x^k} dx^k,\qquad \theta_{i\,\alpha} = dp_{i\, \alpha} - p_{i\, \alpha + x^k} dx^k.
\end{align*}

Let us consider the system
\begin{align}
\begin{aligned}
&l_F^{\hspace{0.15ex} *}(p) = 0\,,\qquad
F = 0\,,
\end{aligned}
\label{cotan}
\end{align}
which is the Euler--Lagrange equation for the Lagrangian
\begin{align}
L = \langle p, F \rangle = p_i\hspace{0.15ex} F^i dx^1\wedge \ldots\wedge dx^n.
\label{CanLag}
\end{align}
Denote by $\mathcal{E}^*$ its infinite prolongation. 
Henceforth, we assume that the system $\mathcal{E}^*$ is $\ell$-normal\footnote{In the case of $\mathcal{E}^*$, all manifolds in a tower of the form~\eqref{ETower} are assumed to be graded.}.
Then $\mathcal{E}$ is also $\ell$-normal and the projection $\hat{\pi}_{\pi}|_{\mathcal{E}^*}\colon \mathcal{E}^*\to \mathcal{E}$ is the \emph{cotangent covering} of~$\mathcal{E}$.

The Lagrangian~\eqref{CanLag} gives rise to the corresponding element of~\eqref{IntSepLag} on $\mathcal{E}^*$ having internal degree $1$. This element is 
defined by the Green formula. More precisely, there exists a Cartan $n$-form $\omega_L\in \mathcal{C}\Lambda^1(\hat\pi)\wedge \hat\pi_{\infty}^{\hspace{0.1ex} *}(\Lambda^{n-1}(M))$ such that for any $\chi\in \varkappa(\pi)$,
\begin{align}
\langle l_F(\chi), p \rangle - \langle \chi, l_F^{\hspace{0.15ex} *}(p) \rangle = d_h (E_{(\chi,\, \ldots)}\lrcorner\, \omega_L)\,.
\label{Green}
\end{align}
We assume, without loss of generality, that $\omega_L$ is linear in $p_{i\,\alpha}$ and does not involve the forms $\theta_{i\,\alpha}$ in adapted coordinates. Then~\eqref{Green} does not depend on the second component of the characteristic $(\chi,\, \ldots)\in \varkappa(\hat{\pi})$. The restriction $\omega_L|_{\mathcal{E}^*}$ determines the corresponding variational $1$-form (of internal degree $1$). We denote it by $\rho$. The canonical variational $1$-form $\rho\in E_1^{1,\, n-1}(\mathcal{E}^*)$ gives rise to the presymplectic structure
\begin{align*}
\Omega = d_1 \rho\in E^{2,\,n-1}_1(\mathcal{E}^*)\,.
\end{align*}
As in the commutative case, the variational $1$-form of $\mathcal{E}^*$ corresponding to the cosymmetry arising from a (graded) symmetry $\Upsilon$ is given by the contraction $\Upsilon \lrcorner\, \Omega$\,. The mapping from cosymmetries of $\mathcal{E}^*$ to its variational $1$-forms (based on~\eqref{Cosymtoonef}, adapted to this case) is an isomorphism.

\vspace{1ex}

\remarka{The cotangent system $\mathcal{E}^*$ gives rise to the variational Schouten bracket and, along with its canonical variational $1$-form, is independent of the choice of $\ell$-normal inclusion $\mathcal{E}\subset J^{\infty}(\pi)$ in the sense of~\cite{KraVer}.}

\subsection{Lifts of symmetries}

Let $X = E_{\varphi}|_{\mathcal{E}}$ be an evolutionary symmetry of $\mathcal{E}$, $\varphi\in\varkappa(\pi)$. The symmetry $X$ can be lifted to the cotangent equation $\mathcal{E}^*$. Namely, there exists a $\mathcal{C}$-differential operator $\Phi\colon P(\pi)\to P(\pi)$ such that
\begin{align}
E_{\varphi}(F) = \Phi(F)\,.
\label{Phi}
\end{align}
Integration by parts in~\eqref{Green} for $\chi = \varphi$ shows that $(-\varphi, \Phi^*(p))$ is a characteristic of a conservation law of the cotangent equation $\mathcal{E}^*$. Then the evolutionary field with the characteristic $(\varphi, -\Phi^*(p))$
is a symmetry of $\mathcal{E}^*$. We denote by $\mathcal{X}$ its restriction to $\mathcal{E}^*$ and say that $\mathcal{X}$ is the \emph{lift} of $X$.

From~\eqref{Green}, one can see that the conservation law with the characteristic $(-\varphi, \Phi^*(p))$ is $\mathcal{X}\lrcorner\, \rho$. 
Then the formula $\mathcal{L}_{\mathcal{X}} \rho = \mathcal{X} \lrcorner\, \Omega + d_1(\mathcal{X}\lrcorner\, \rho)$ and a simple analysis of cosymmetries lead to the following important result.

\vspace{1ex}

\lemmaa{\label{lemma1} Let $X = E_{\varphi}|_{\mathcal{E}}$ be a symmetry of $\mathcal{E}$. Then for its lift $\mathcal{X}$, the formula
\begin{align}
\mathcal{L}_{\mathcal{X}} \rho = 0
\label{Canoninv}
\end{align}
holds. In other words, the canonical variational $1$-form $\rho$ is $\mathcal{X}$-invariant.
}

\section{\label{HamopSect} Hamiltonian operators}

Let us recall basic facts about Hamiltonian operators of $\ell$-normal PDEs and describe them in terms of conservation laws and symmetries of the cotangent equations.

Let $\nabla\colon \widehat{P}(\mathcal{E})\to \varkappa(\mathcal{E})$ be a $\mathcal{C}$-differential operator such that
\begin{align}
l_{\mathcal{E}}\circ \nabla - \nabla^*\circ l_{\mathcal{E}}^{\hspace{0.15ex} *} = 0\,.
\label{biveconsys}
\end{align}
Denote by $\nabla_{e}$ some extension of $\nabla$ to $J^{\infty}(\pi)$, i.e., a $\mathcal{C}$-differential operator $\nabla_e\colon \widehat{P}(\pi)\to \varkappa(\pi)$ such that $\nabla = \nabla_e|_{\mathcal{E}}$. Then there exists a $\mathcal{C}$-differential operator $\Delta\colon P(\pi)\times \widehat{P}(\pi)\to P(\pi)$ satisfying
\begin{align}
l_F\circ \nabla_e - \nabla_e^{\,*}\circ l_F^{\hspace{0.15ex} *} = \Delta(F, \cdot)\,.
\label{bivec}
\end{align}
It is convenient to denote $\Delta(\cdot, \psi)$ by $\Delta_{\psi}(\cdot)$.

Operators satisfying~\eqref{biveconsys} define skew-symmetric brackets on the vector space of conservation laws by means of the Lie derivative. Namely, if $\xi_1, \xi_2$ are conservation laws of $\mathcal{E}$ and $\,\overline{\!\psi}_1\in \ker l_{\mathcal{E}}^{\hspace{0.15ex} *}$ is the cosymmetry of $\xi_1$, then $E_{\nabla(\,\overline{\!\psi}_1)} = E_{\nabla_e(\psi_1)}|_{\mathcal{E}}$ is a symmetry. The corresponding bracket is
\begin{align}
\{\xi_1, \xi_2\}_{\nabla} = \mathcal{L}_{E_{\nabla(\,\overline{\!\psi}_1)}}\xi_2\,.
\label{Nablabracket}
\end{align}

\remarka{\label{remB} Using~\eqref{bivec}, one can show that the operator
 $B\colon \widehat{P}(\pi)\times \widehat{P}(\pi)\to \widehat{P}(\pi)$,
\begin{align*}
B(\psi_1, \psi_2) = \Delta_{\psi_1}^*(\psi_2) + \Delta_{\psi_2}^*(\psi_1)
\end{align*}
yields a characteristic of a conservation law of $\mathcal{E}$ for any $\psi_1, \psi_2\in \widehat{P}(\pi)$. Then $l_{\mathcal{E}}^{\hspace{0.15ex} *}\circ B|_{\mathcal{E}} = 0$. Since $\mathcal{E}$ is $\ell$-normal, this condition implies $B|_{\mathcal{E}} = 0$.}

\subsection{\label{Hamopvarf} Variational bivectors and variational $1$-forms}

Operators satisfying~\eqref{biveconsys} represent variational bivectors~\cite{KraVer, KerKraVerVit}.
They map cosymmetries of $\mathcal{E}$ to its symmetries. Using the isomorphism between $E^{1,\,n-1}_1(\mathcal{E})$ and $\ker l_{\mathcal{E}}^{\hspace{0.15ex} *}$, one can interpret them as mappings from variational $1$-forms of $\mathcal{E}$ to its symmetries. For a variational $1$-form $\nu\in E^{1,\,n-1}_1(\mathcal{E})$, we denote by $X_{\nabla(\nu)}$ the respective symmetry of $\mathcal{E}$.

Let $\nu$ be a variational $1$-form of $\mathcal{E}$. There exist $\omega_{\nu}\in \mathcal{C}\Lambda^1(\pi)\wedge\pi_{\infty}^{\hspace{0.1ex} *}(\Lambda^{n-1}(M))$ and $\psi_{\nu}\in \widehat{P}(\pi)$ such that $\psi_{\nu}|_{\mathcal{E}}$ is the cosymmetry corresponding to $\nu$, the restriction $\omega_{\nu}|_{\mathcal{E}}$ represents $\nu\in E^{1,\,n-1}_1(\mathcal{E})$, and for any $\chi\in \varkappa(\pi)$,
\begin{align*}
\langle l_F(\chi), \psi_{\nu} \rangle = \langle \chi, l_F^{\hspace{0.15ex} *}(\psi_{\nu}) \rangle + d_h (E_{\chi}\lrcorner\, \omega_{\nu})\,.
\end{align*}
In addition, there exists a $\mathcal{C}$-differential operator $\Psi\colon P(\pi)\to \widehat{\varkappa}(\pi)$ such that
\begin{align*}
l_F^{\hspace{0.15ex} *}(\psi_{\nu}) = \Psi(F)\,.
\end{align*}

\lemmaa{\label{lemma2} The lift $\mathcal{X}_{\nabla(\nu)}$ of the symmetry $X_{\nabla(\nu)}$ is the restriction of the evolutionary vector field with the characteristic $(\nabla_e(\psi_{\nu}), \Delta_p^*(\psi_{\nu}) - \Psi^*\nabla_e(p))$ to $\mathcal{E}^*$.
}

\vspace{0.7ex}

\noindent
\textbf{Proof.} We need to find $\Phi$ from~\eqref{Phi}. Using~\eqref{bivec}, one obtains
\begin{align*}
E_{\nabla_e(\psi_\nu)}(F) = l_F \nabla_e(\psi_\nu) = \nabla_e^{\hspace{0.15ex} *} l_F^{\hspace{0.15ex} *} (\psi_\nu) + \Delta_{\psi_{\nu}}(F) = \nabla_e^{\hspace{0.15ex} *} \Psi(F) + \Delta_{\psi_{\nu}}(F)\,.
\end{align*}
Then one can take $\Phi = \nabla_e^{\hspace{0.15ex} *} \Psi + \Delta_{\psi_{\nu}}$. Hence, $\Phi^*(p) = \Psi^* \nabla_e (p) + \Delta_{\psi_{\nu}}^*(p)$, and the symmetry $\mathcal{X}_{\nabla(\nu)}$ is the restriction of the evolutionary vector field with the characteristic
\begin{align*}
(\nabla_e(\psi_\nu), -\Delta_{\psi_{\nu}}^*(p) - \Psi^* \nabla_e (p))\,.
\end{align*}
The observation from Remark~\ref{remB} completes the proof.

\subsection{Variational bivectors as odd symmetries}

One can generalize~\eqref{Green} and take $\nabla_e(p)$ instead of $\chi$, while keeping track of signs. Then
\begin{align}
\langle l_F\nabla_e(p), p \rangle - \langle \nabla_e(p), l_F^{\hspace{0.15ex} *}(p) \rangle = d_h (E_{(\nabla_e(p),\, \ldots)}\,\lrcorner\, \omega_L)\,.
\label{genGreen}
\end{align}
Using~\eqref{bivec}, one obtains
\begin{align}
\langle \nabla_e^{\,*}l_F^{\hspace{0.15ex} *}(p), p \rangle + \langle \Delta_p(F), p \rangle - \langle \nabla_e(p), l_F^{\hspace{0.15ex} *}(p) \rangle = d_h (E_{(\nabla_e(p),\, \ldots)}\,\lrcorner\, \omega_L)\,.
\label{intbyparnab}
\end{align}
Then
$
(-2\nabla_e(p), \Delta_p^{*}(p))
$
is a characteristic of a conservation law of $\mathcal{E}^*$. Hence, the restriction $s_{\nabla}$ of the evolutionary vector field with the characteristic $(\nabla_e(p), -\frac{1}{2}\Delta_p^{*}(p))$
is a degree-$1$ symmetry of $\mathcal{E}^*$. Let us stress that the odd parity (degree mod $2$) of the variables $p_{i\, \alpha}$ plays a crucial role here.

From~\eqref{intbyparnab}, one finds that
\begin{align}
s_{\nabla}\lrcorner\, d_1 \rho = d_1 \mathcal{H}_{\nabla}\,,
\label{sNoeth}
\end{align}
where $\mathcal{H}_{\nabla}$ denotes the conservation law of internal degree $2$
\begin{align}
\mathcal{H}_{\nabla} = -\dfrac{1}{2} s_{\nabla} \lrcorner\, \rho\,.
\label{Hamil}
\end{align}
Note that $\mathcal{H}_{\nabla}$ depends only\footnote{Similarly, $\mathcal{X}\lrcorner\, \rho$ (and hence $\mathcal{X}$) depends only on $X$. Then one can show that the conditions $\mathcal{X}|_{\mathcal{F}(\mathcal{E})} = X$ and $\mathcal{L}_{\mathcal{X}}\rho = 0$ together completely determine $\mathcal{X}$. Using this, one can show that the lift preserves commutators.} on $\nabla$. Moreover, any $\mathcal{C}$-differential operator of the form $\nabla + \square\circ l_{\mathcal{E}}^{\hspace{0.15ex} *}$ leads to the same conservation law $\mathcal{H}_{\nabla}$, where $\square\colon \widehat{\varkappa}(\mathcal{E})\to \varkappa(\mathcal{E})$ and $\square^* = \square$. Then the equivalence class $\nabla + \square\circ l_{\mathcal{E}}^{\hspace{0.15ex} *}$ (i.e., the corresponding variational bivector) determines the same symmetry $s_{\nabla}$.

\vspace{1ex}

\remarka{The identification of the mutually corresponding variational bivectors arising from different $\ell$-normal inclusions of $\mathcal{E}$ into jet spaces (assuming that the arising $\mathcal{E}^*$ are also $\ell$-normal) allows us to conclude that the symmetry $s_{\nabla}$ and the subsequent reduction mechanism are independent of the choice of such an inclusion, thereby ensuring the canonical nature of our approach.}

\vspace{1ex}

\definitiona{An operator $\nabla$ satisfying~\eqref{biveconsys} is a \textit{Hamiltonian operator} of $\mathcal{E}$ if $s_{\nabla}$ is a cohomological vector field, i.e.,
\begin{align*}
[s_{\nabla}, s_{\nabla}] = 0\,.
\end{align*}}

\noindent
Here $[s_{\nabla}, s_{\nabla}] = 2s_{\nabla}\circ s_{\nabla}$ since $s_{\nabla}$ is odd. If $\nabla$ is Hamiltonian, the bracket~\eqref{Nablabracket} enjoys the Jacobi identity.

If $X = E_{\varphi}|_{\mathcal{E}}$ is a symmetry of $\mathcal{E}$, we say that $\nabla$ is $X$-invariant (or that $X$ preserves $\{\cdot, \cdot\}_{\nabla}$) if the symmetry $s_{\nabla}$ commutes with the lift of $X$,
\begin{align*}
[s_{\nabla}, \mathcal{X}] = 0\,.
\end{align*}
This condition is equivalent\footnote{This condition also implies that $s_{\nabla}$ can be restricted to $\mathcal{E}^*_{\mathcal{X}}$.} to the $\mathcal{X}$-invariance of the conservation law $\mathcal{H}_{\nabla}$, i.e., to $\mathcal{L}_{\mathcal{X}} \mathcal{H}_{\nabla} = 0$.

\vspace{1ex}

\remarka{It was shown in~\cite{KerKraVer} that, for evolution systems satisfying a certain non-degeneracy condition, every operator $\nabla$ satisfying~\eqref{biveconsys} is Hamiltonian.
}

\vspace{1ex}

\remarka{Hamiltonian operators $\nabla_1$ and $\nabla_2$ form a Poisson pencil if $[s_{\nabla_1}, s_{\nabla_2}] = 0$.
}

\subsection{Action of variational bivectors in terms of odd symmetries}

Let $\nu\in E^{1,\,n-1}_1(\mathcal{E})$ be a variational $1$-form, $\psi_\nu$, $\omega_\nu$, and $\Psi$ be the corresponding structures from Section~\ref{Hamopvarf}.
Then for any $\chi\in \varkappa(\pi)$,
\begin{align*}
\langle l_F(\chi), \psi_{\nu} \rangle = \langle \chi, l_F^{\hspace{0.15ex} *}(\psi_{\nu}) \rangle + d_h (E_{(\chi,\, \ldots)}\lrcorner\, \hat{\pi}_{\pi}^{\hspace{0.1ex} *}(\omega_{\nu}))\,.
\end{align*}
Replacing $\chi$ by $\nabla_e(p)$ and using~\eqref{bivec}, one finds
\begin{align*}
\langle \nabla_e^{*}\,l_F^{\hspace{0.15ex} *}(p) + \Delta_p(F), \psi_{\nu} \rangle - \langle \nabla_e(p), \Psi(F) \rangle = d_h (E_{(\nabla_e(p),\, \ldots)}\lrcorner\, \hat{\pi}_{\pi}^{\hspace{0.1ex} *}(\omega_{\nu}))\,.
\end{align*}
Therefore, the conservation law with the characteristic
\begin{align*}
(\nabla_e(\psi_{\nu}), \Delta_p^*(\psi_{\nu}) - \Psi^*\nabla_e(p))
\end{align*}
is $s_{\nabla}\lrcorner\, {\nu}$, where ${\nu}$ is treated as an element of $E^{1, \hspace{0.1ex} n-1}_1(\mathcal{E}^*)$ due to the inclusion $E^{1, \hspace{0.1ex} n-1}_1(\mathcal{E}) \subset E^{1, \hspace{0.1ex} n-1}_1(\mathcal{E}^*)$ defined by the cotangent covering. According to Lemma~\ref{lemma2}, the following formula holds
\begin{align}
d_1 (s_{\nabla}\lrcorner\, {\nu}) = \mathcal{X}_{\nabla(\nu)}\lrcorner\, \Omega\,.
\label{actionnablagenomeg}
\end{align}
It describes the action of $\nabla$ on variational $1$-forms of $\mathcal{E}$ in terms of $s_{\nabla}$, i.e., it can be used as an equivalent definition of 
$X_{\nabla(\nu)}$. Moreover, the identity $\ker d_1^{\hspace{0.1ex} 0, \hspace{0.1ex} n-1} = 0$ for $\mathcal{E}^*$ and Lemma~\ref{lemma1} imply
\begin{align}
s_{\nabla}\lrcorner\, {\nu} = -\mathcal{X}_{\nabla(\nu)}\lrcorner\, \rho\,.
\label{actionnablagen}
\end{align}
The importance of this formula lies in the fact that, unlike $\nabla$, both its ingredients and the formula itself admit the invariant reduction.

In particular, if $\xi\in E^{0, \hspace{0.1ex} n-1}_1(\mathcal{E}) \subset E^{0, \hspace{0.1ex} n-1}_1(\mathcal{E}^*)$ is a conservation law of $\mathcal{E}$, the corresponding symmetry $\mathcal{X}_{\nabla(d_1\xi)}$ satisfies
\begin{align}
s_{\nabla}\lrcorner\, d_1 {\xi} = -\mathcal{X}_{\nabla(d_1\xi)}\lrcorner\, \rho\,.
\label{nablaass}
\end{align}

\remarka{If $\xi$ is a conservation law of $\mathcal{E}$, then there is the evolutionary symmetry $\Xi$
of $\mathcal{E}^*$ such that $\Xi\, \lrcorner\, \Omega = d_1{\xi}$. The symmetry $\Xi$ has degree $-1$. Applying the Lie derivative $\mathcal{L}_{s_{\nabla}}$ to the relation $\Xi\, \lrcorner\, \Omega = d_1{\xi}$, one can show that $[s_{\nabla}, \Xi] = \mathcal{X}_{\nabla(d_1\xi)}$. It follows that if $\nabla$ is a Hamiltonian operator, then the Hamiltonian vector field $X_{\nabla(d_1\xi)}$ preserves\footnote{It also follows that $[[s_{\nabla}, \Xi_1], \Xi_2]\lrcorner\, \Omega = d_1\{\xi_1, \xi_2\}_{\nabla}$, where $\Xi_i\, \lrcorner\, \Omega = d_1{\xi_i}$ for $\xi_i\in E^{0, \hspace{0.1ex} n-1}_1(\mathcal{E}) \subset E^{0, \hspace{0.1ex} n-1}_1(\mathcal{E}^*)$. This relation implies that the bracket~\eqref{Nablabracket} is skew-symmetric (and that it enjoys the Jacobi identity if $\nabla$ is Hamiltonian).} $\{\cdot, \cdot\}_{\nabla}$~(see \cite{KraVer}). Thus, for any conservation law $\xi\in E^{0, \hspace{0.1ex} n-1}_1(\mathcal{E})$, a Hamiltonian operator $\nabla$ is $X_{\nabla(d_1\xi)}$-invariant.}

\vspace{1ex}

\remarka{If $X$ is an evolutionary symmetry of $\mathcal{E}$ such that both $\nabla$ and $\nu$ in~\eqref{actionnablagenomeg} are $X$-invariant, then applying the Lie derivative $\mathcal{L}_{\mathcal{X}}$ to~\eqref{actionnablagenomeg}, one obtains $[\mathcal{X}, \mathcal{X}_{\nabla(\nu)}] = 0$ and hence $[X, X_{\nabla(\nu)}] = 0$.}

\section{\label{invredmechsec} Invariant reduction mechanism}

Let us recall the invariant reduction mechanism~\cite{DrCh2} and apply it in both commutative and graded-commutative cases.

Let $X$ be an evolutionary symmetry of $\mathcal{E}\subset J^{\infty}(\pi)$. Denote by $\mathcal{E}_X$ the subsystem\footnote{$\mathcal{E}_X\subset \mathcal{E}$ is characterized
by the condition that $X$ vanishes at its points.} given by the infinite prolongation of the system
\begin{align*}
&F = 0\,,\qquad
\varphi = 0\,,
\end{align*}
where $X = E_{\varphi}|_{\mathcal{E}}$, while $F = 0$ determines $\mathcal{E}$.
Suppose $\omega\in E_0^{p, \hspace{0.1ex} q}(\mathcal{E})$ represents an $X$-invariant element of $E_1^{p, \hspace{0.1ex} q}(\mathcal{E})$, where $q \geqslant 1$. Then there exists $\vartheta\in E_0^{p, \hspace{0.1ex} q-1}(\mathcal{E})$ such that
\begin{align*}
\mathcal{L}_X\omega = d_0 \vartheta.
\end{align*}
Restricting this relation to $\mathcal{E}_X$, we obtain
\begin{align*}
d_0 \vartheta|_{\mathcal{E}_X} = 0\,.
\end{align*}
The invariant reduction of the $X$-invariant element represented by $\omega$ is the element of $E_1^{p, \hspace{0.1ex} q-1}(\mathcal{E}_X)$ represented by $\vartheta|_{\mathcal{E}_X}$. We denote by $\mathcal{R}_X^{p, \hspace{0.1ex} q}$ the homomorphism that evaluates the invariant reduction of $X$-invariant elements of $E_1^{p, \hspace{0.1ex} q}(\mathcal{E})$, provided it is well-defined. This is the case when $E_1^{p, \hspace{0.1ex} q-1}(\mathcal{E})|_{\mathcal{E}_X} =~\!0$. If $E_1^{0, \hspace{0.1ex} 0}(\mathcal{E})|_{\mathcal{E}_X} \subset H^0_{dR}(\mathcal{E}_X)$, we also say that the invariant reduction of $X$-invariant elements of $E_1^{0, \hspace{0.1ex} 1}(\mathcal{E})$ is well-defined, as we consider their reductions up to additive locally constant functions on $\mathcal{E}_X$.
The graded-commutative counterpart arises from the replacement of $(\mathcal{E}, X)$ by $(\mathcal{E}^*, \mathcal{X})$, where $\mathcal{X}$ is the lift of $X$. In this case, we additionally assume, without loss of generality, that $\omega$ and $\vartheta$ have the same internal degree.

The following two theorems are given in~\cite{DrCh2} (as Theorems 1 and~2). Their formulations (and the proofs from~\cite{DrCh2}) carry over verbatim to the graded-commutative case upon replacing $(\mathcal{E}, X)$ with $(\mathcal{E}^*, \mathcal{X})$, where $\mathcal{X}$ is the lift of $X$, and assuming that some internal degree is fixed.

\vspace{1ex}

\theorema{\label{Theor1} Let $\mathcal{E}$ be an infinitely prolonged system of differential equations, and let $X = E_{\varphi}|_{\mathcal{E}}$ be its symmetry.
Suppose that the invariant reduction is well-defined for $X$-invariant elements of $E_1^{p, \hspace{0.1ex} q}(\mathcal{E})$ and $E_1^{p+1, \hspace{0.1ex} q}(\mathcal{E})$. Then on the $X$-invariant subspace of~$E_1^{p, \hspace{0.1ex} q}(\mathcal{E})$,}
$$
\mathcal{R}_X^{p+1, \hspace{0.1ex} q}\circ d_1 = - d_1\circ\mathcal{R}_X^{p, \hspace{0.1ex} q}
$$

\theorema{\label{Theor2} Suppose that $X = E_{\varphi}|_{\mathcal{E}}$, $X_1 = E_{\varphi_1}|_{\mathcal{E}}$ are commuting symmetries of an infinitely prolonged system $\mathcal{E}$. If the invariant reduction is well-defined for $X$-invariant elements of $E_1^{p, \hspace{0.1ex} q}(\mathcal{E})$ and $E_1^{p-1, \hspace{0.1ex} q}(\mathcal{E})$, then on the $X$-invariant subspace of\,\footnote{If $(p; q) = (1; 1)$, the relation~\eqref{Theor2relation} is understood up to an additive locally constant function on $\mathcal{E}_X$.} $E_1^{p, \hspace{0.1ex} q}(\mathcal{E})$,}
\begin{align}\label{Theor2relation}
\mathcal{R}_X^{p-1, \hspace{0.1ex} q}\circ {X_1} \lrcorner = -X_1|_{\mathcal{E}_X} \lrcorner\, \circ \mathcal{R}_X^{p, \hspace{0.1ex} q}
\end{align}

\noindent
Below we omit the superscripts of the invariant reduction homomorphisms and apply the reduction to invariant elements of groups $E_1^{\ast, \hspace{0.1ex} n-1}$ of the $\ell$-normal systems $\mathcal{E}$ and $\mathcal{E}^*$. In this case, the invariant reduction is well-defined. In particular, using Lemma~\ref{lemma1}, one obtains
\begin{align*}
\mathcal{R}_{\mathcal{X}}(\rho)\in E^{1, \hspace{0.1ex} n-2}_1(\mathcal{E}^*_{\mathcal{X}})\,.
\end{align*}
Here $\mathcal{X}$ is the lift of a symmetry $X = E_{\varphi}|_{\mathcal{E}}$ of $\mathcal{E}$, the system $\mathcal{E}^*_{\mathcal{X}}$ is the infinite prolongation of
\begin{align*}
l_F^{\hspace{0.15ex} *}(p) = 0\,,\qquad F = 0\,,\qquad \varphi = 0\,,\qquad \Phi^*(p) = 0
\end{align*}
for any operator $\Phi$ from~\eqref{Phi}.

\vspace{1ex}

\remarka{Elements of the groups $E_1^{p,\hspace{0.1ex} 0}$, $p\geqslant 1$ are $d_0$-closed differential forms (not equivalence classes).
The differential $d_1^{\hspace{0.1ex} *, \hspace{0.1ex} 0}\colon E_1^{*,\hspace{0.1ex} 0}\to E_1^{*,\hspace{0.1ex} 0}$ coincides with the de Rham differential $d$.}

\section{\label{finitedimcovsec} Cotangent coverings of finite-dimensional systems}

For finite-dimensional differential equations, we can intrinsically introduce objects that play the role of cotangent equations. For simplicity, we restrict ourselves to surjective submersions.

Let $\pi_{\mathcal{S}}\colon \mathcal{S}^{n+N}\to M^n$ be a smooth surjective submersion between finite-dimensional connected smooth manifolds. We assume that it is equipped with a structure of a differential equation, i.e., a flat connection\footnote{It is a differential equation in a sense that is more general than the one adopted in Section~\ref{SectionDiffEq}. The connection canonically defines an embedding of $\mathcal{S}$ into the manifold $J^1(\pi_{\mathcal{S}})$ of $1$-jets of local sections of the submersion~$\pi_{\mathcal{S}}$. We do not require the connection to have global path lifting.}. Denote by $D^v(\mathcal{S})$ the module of $\pi_{\mathcal{S}}$-vertical derivations on $\mathcal{S}$. The decomposition
\begin{align*}
D(\mathcal{S}) = \mathcal{C}D(\mathcal{S})\oplus D^v(\mathcal{S})
\end{align*}
of the module $D(\mathcal{S})$ of derivations on $\mathcal{S}$ allows one to define an action of Cartan derivations on $D^v(\mathcal{S})$ using the vertical components of the corresponding Lie derivatives. The finitely generated projective module $D^v(\mathcal{S})$ can be identified with the module of fiberwise linear functions on a smooth vector bundle $\tau^*_{\mathcal{S}}\colon \mathfrak{T}^*\mathcal{S}\to \mathcal{S}$ (see~\cite{Nestruev}).
Since Cartan derivations act on $D^v(\mathcal{S})$, the bundle $\tau^*_{\mathcal{S}}\circ \pi_{\mathcal{S}} \colon \mathfrak{T}^*\mathcal{S}\to M$ is endowed with the corresponding flat connection. In other words, $\mathfrak{T}^*\mathcal{S}$ is a differential equation. Note that $\tau^*_\mathcal{S}$ is a differential covering.

Denote by $\mathfrak{T}^*[1]\mathcal{S}$ the degree-shifted version of $\mathfrak{T}^*\mathcal{S}$, whose algebra of differential forms is also bigraded.
Cartan derivations act on the module $\mathcal{C}\Lambda^1(\mathcal{S})$ by means of the Lie derivative. Then they act on $\mathcal{C}\Lambda^1(\mathcal{S}) \otimes D^v(\mathcal{S})$ by means of the Leibniz rule. In terms of the isomorphism\footnote{The tensor product $\mathcal{C}\Lambda^1(\mathcal{S}) \otimes D^v(\mathcal{S})$ is taken over the algebra $\mathcal{F}(\mathcal{S})$, and $\mathrm{End}(D^v(\mathcal{S}))$ denotes $\mathrm{End}_{\mathcal{F}(\mathcal{S})}(D^v(\mathcal{S}))$.}
\begin{align*}
\mathcal{C}\Lambda^1(\mathcal{S}) \otimes D^v(\mathcal{S}) = \mathrm{End}(D^v(\mathcal{S}))\,,
\end{align*}
this action can be described as follows. For any $\mu\in \mathrm{End}(D^v(\mathcal{S}))$, $w\in D^v(\mathcal{S})$, and $Y\in \mathcal{C}D(\mathcal{S})$,
\begin{align}
Y(\mu)\, w = Y(\mu\, w) - \mu\, Y(w)\,.
\label{Leibnizru}
\end{align}
Here $Y(w)$ is the $\pi_{\mathcal{S}}$-vertical component of $\mathcal{L}_{Y}w = [Y, w]$. Note that if $Y$ is the horizontal lift of a vector field from $M$, then $Y(w) = [Y, w]$ for any vertical vector field $w$.

Let us introduce local coordinates $x = (x^1, \ldots, x^n)$ on $M$ and $(x, v) = (x^1, \ldots, x^n, v^1, \ldots, v^N)$ on $\mathcal{S}$ such that in these coordinates, $\pi_{\mathcal{S}}$ takes the form $\pi_{\mathcal{S}}(x, v) = x$. The basis of local Cartan $1$-forms and the total derivatives are given by
\begin{align*}
\gamma^i = dv^i - \Gamma^i_k dx^k\,,\qquad \widetilde{D}_{x^k} = \partial_{x^k} + \Gamma^i_k\partial_{v^i}
\end{align*}
for some functions $\Gamma^i_k(x, v)$. Denote $\varpi_j = \partial_{v^j}$. Then, in the corresponding neighborhood on
$\mathfrak{T}^*[1]\mathcal{S}$, the basis of Cartan $1$-forms and the total derivatives take the form
\begin{align*}
\gamma^{j} = dv^j - \Gamma^j_k dx^k,\qquad \gamma_j = d\varpi_j + \dfrac{\partial \Gamma^i_k}{\partial v^j} \varpi_i\, dx^k,\qquad \mathfrak{D}_{x^k} = \partial_{x^k} + \Gamma^i_k\partial_{v^i} - \dfrac{\partial \Gamma^i_k}{\partial v^j} \varpi_i \partial_{\varpi_j}.
\end{align*}

The identity operator $\mathbb{I}\in \mathrm{End}(D^v(\mathcal{S}))$
defines the Cartan $1$-form $\rho_c = \gamma^{j} \varpi_j \in \mathcal{C}\Lambda^1(\mathfrak{T}^*[1]{\mathcal{S}})$. 
It follows from~\eqref{Leibnizru} that $Y(\mathbb{I}) = 0$ for any $Y\in \mathcal{C}D(\mathcal{S})$. This is equivalent to the condition that $d_0 \rho_c = 0$, which in turn implies
\begin{align*}
d\rho_c = \gamma_j \wedge \gamma^{j}.
\end{align*}

\remarka{\label{remker}
The kernel distribution of $d\rho_c\in E^{2, \hspace{0.1ex} 0}_1(\mathfrak{T}^*[1]\mathcal{S})$ is the Cartan distribution of $\mathfrak{T}^*[1]\mathcal{S}$.
}

\vspace{1ex}

\remarka{ \label{Remfibers} Using the corresponding form $\rho_c\in E^{1, \hspace{0.1ex} 0}_1(\mathfrak{T}^*\mathcal{S})$, one can show that
for any $\beta\in \mathcal{C}\Lambda^1(\mathcal{S})$ and any point of $\mathcal{S}$, there exists a unique point in the corresponding fiber of $\tau^*_{\mathcal{S}}$ at which $\rho_c$ coincides with the pullback of $\beta$. This reflects the fact that the module of sections of the bundle $\tau^*_\mathcal{S}$ is $\mathcal{C}\Lambda^1(\mathcal{S})$.
}

\vspace{1ex}

Let us adapt the symbol $\simeq$ for canonically related objects on $\mathcal{S}$ and $\mathfrak{T}^*[1]\mathcal{S}$. In particular,
$$
\mathbb{I} \simeq \rho_c\in E^{1, \hspace{0.1ex} 0}_1(\mathfrak{T}^*[1]\mathcal{S})\,.
$$
Consider a $\pi_{\mathcal{S}}$-vertical bivector $b = b^{ij}(x, v)\,\partial_{v^i} \wedge \partial_{v^j}$ on $\mathcal{S}$. Then for the function $H_b = b^{ij}\varpi_i \varpi_j \simeq b$, we have the following

\vspace{1ex}

\propositiona{\label{theorem3} If $H_b \in E^{0, \hspace{0.1ex} 0}_1(\mathfrak{T}^*[1]\mathcal{S})$, then $b(\lambda, \cdot)$ is a symmetry of $\mathcal{S}$ for any 
$\lambda\in E^{1, \hspace{0.1ex} 0}_1(\mathcal{S})$.
}
\vspace{0.5ex}

\noindent
\textbf{Proof.} The conditions $d_0 H_b = 0$ and $d_0\lambda = 0$ can be rewritten as follows. For any $Y\in \mathcal{C}D(\mathcal{S})$,
\begin{align*}
\mathcal{L}_Y b \in \mathcal{C}D(\mathcal{S})\wedge D^v(\mathcal{S})\,,\qquad \mathcal{L}_Y \lambda = 0\,.
\end{align*}
Consequently, for any $Y\in \mathcal{C}D(\mathcal{S})$,
$$
[b(\lambda, \cdot), Y] = -\mathcal{L}_Y (b(\lambda, \cdot)) = -(\mathcal{L}_Y b)(\lambda, \cdot)\in \mathcal{C}D(\mathcal{S})\,,
$$
i.e., the vector field $b(\lambda, \cdot)$ preserves the module $\mathcal{C}D(\mathcal{S})$. This observation completes the proof.

\vspace{1ex}

Let $H_b \in E^{0, \hspace{0.1ex} 0}_1(\mathfrak{T}^*[1]\mathcal{S})$. Then $b$ gives rise to the degree-$1$ symmetry $s_b$ of $\mathfrak{T}^*[1]\mathcal{S}$ such that
\begin{align}
s_b\hspace{0.15ex} \lrcorner\, d\rho_c = -dH_b\,.
\label{Finitedimrel}
\end{align}
In coordinates, it has the form
\begin{align*}
s_b = -(b^{ij} - b^{ji})\varpi_j\partial_{v^i} - \dfrac{\partial b^{ij}}{\partial v^k}\varpi_{i}\varpi_j \partial_{\varpi_k}\,.
\end{align*}

\remarka{For a Cartan $1$-form $\beta\in \mathcal{C}\Lambda^1(\mathcal{S})\subset \mathcal{C}\Lambda^1(\mathfrak{T}^*[1]\mathcal{S})$, one has
\begin{align}
b(\beta, \cdot) \simeq - s_b\, \lrcorner\, \beta\,.
\label{bracketmech}
\end{align}
}

The following proposition arises from the characterization of Poisson bivectors on fibers of $\pi_{\mathcal{S}}$ via the relation between the Schouten bracket and the graded commutator.

\vspace{1ex}

\propositiona{\label{theorem4} If $H_b \in E^{0, \hspace{0.1ex} 0}_1(\mathfrak{T}^*[1]\mathcal{S})$, then $b$ is a Poisson bivector if and only if $[s_b, s_b] = 0$.
}

\vspace{1ex}

\noindent
In particular, Propositions~\ref{theorem3} and~\ref{theorem4} show that if $H_b \in E^{0, \hspace{0.1ex} 0}_1(\mathfrak{T}^*[1]\mathcal{S})$ and $[s_b, s_b] = 0$, then 
$$
\{h_1, h_2\} = b(dh_1, dh_2)
$$
is a Poisson bracket on the algebra $E^{\hspace{0.1ex} 0, \hspace{0.2ex} 0}_1(\mathcal{S})$ of constants of motion. It is also a Poisson bracket on the algebra of functions~$\mathcal{F}(\mathcal{S})$, although this fact is apparently not of major importance, as the bracket on~$\mathcal{F}(\mathcal{S})$ depends not only on the Cartan distribution of $\mathcal{S}$, but also on the submersion~$\pi_{\mathcal{S}}$.

\section{\label{finitedimredsec} Finite-dimensional reductions}

Let us consider the case $n=2$. 
Suppose that for a symmetry $X = E_{\varphi}|_{\mathcal{E}}$ of $\mathcal{E}$, the system $\mathcal{E}_X$ is a finite-dimensional connected smooth manifold and there is an isomorphism of the bundles $\mathfrak{T}^*[1]\mathcal{E}_X$ and $\mathcal{E}^*_{\mathcal{X}}$ over $\mathcal{E}_X$ relating the differential forms $\rho_c$ and $\mathcal{R}_{\mathcal{X}}(\rho)$. Note that such an isomorphism is unique. As Remark~\ref{remker} and a simple dimension count show, the isomorphism relates the Cartan distributions of $\mathfrak{T}^*[1]\mathcal{E}_X$ and $\mathcal{E}^*_{\mathcal{X}}$ as the kernel distributions of $d\rho_c$ and $d\hspace{0.1ex} \mathcal{R}_{\mathcal{X}}(\rho)$, respectively.

Suppose the system $\mathcal{E}$ admits an $X$-invariant Hamiltonian operator $\nabla$. Applying Theorems~\ref{Theor1} and~\ref{Theor2} (adapted to the graded-commutative case), one sees that the reduction of~\eqref{sNoeth} takes the~form
\begin{align*}
s_{\nabla}|_{\mathcal{E}^*_{\mathcal{X}}}\,  \lrcorner\ d\hspace{0.1ex} \mathcal{R}_{\mathcal{X}}(\rho) = - d\hspace{0.1ex} \mathcal{R}_{\mathcal{X}}(\mathcal{H}_{\nabla})\,,
\end{align*}
which matches the formula~\eqref{Finitedimrel} for $\mathfrak{T}^*[1]\mathcal{E}_X$. Here $\mathcal{R}_{\mathcal{X}}(\mathcal{H}_{\nabla})\in E^{0, \hspace{0.1ex} 0}_1(\mathcal{E}^*_{\mathcal{X}})$ has degree $2$. It is unambiguously defined. In addition, we have
\begin{align*}
[s_{\nabla}|_{\mathcal{E}^*_{\mathcal{X}}}, s_{\nabla}|_{\mathcal{E}^*_{\mathcal{X}}}] = [s_{\nabla}, s_{\nabla}]|_{\mathcal{E}^*_{\mathcal{X}}} = 0\,.
\end{align*}
According to Propositions~\ref{theorem3} and~\ref{theorem4}, the bivector corresponding to $\mathcal{R}_{\mathcal{X}}(\mathcal{H}_{\nabla})$ defines a Poisson bracket on the algebra $E^{0, \hspace{0.1ex} 0}_1(\mathcal{E}_X)$. The action of the bivector on Cartan $1$-forms is defined by~\eqref{bracketmech}. Note that, due to the reduction of~\eqref{Hamil}, direct computation of $\mathcal{R}_{\mathcal{X}}(\mathcal{H}_{\nabla})$ is unnecessary.

For an $X$-invariant conservation law $\xi\in E^{0,\hspace{0.1ex} n-1}_1(\mathcal{E}) \subset E^{0,\hspace{0.1ex} n-1}_1(\mathcal{E}^*)$, one has $\mathcal{R}_{\mathcal{X}}({\xi}) = \mathcal{R}_{X}({\xi})$ due to the inclusion $\mathcal{F}(\mathcal{E}_X)\subset \mathcal{F}(\mathcal{E}^*_{\mathcal{X}})$ induced by the projection $\hat{\pi}_{\pi}|_{\mathcal{E}^*_{\mathcal{X}}}$, and the reduction of~\eqref{nablaass}~yields
\begin{align*}
s_{\nabla}|_{\mathcal{E}^*_{\mathcal{X}}}\, \lrcorner\ d\hspace{0.1ex} \mathcal{R}_{{X}}({\xi}) = \mathcal{X}_{\nabla(d_1\xi)}|_{\mathcal{E}^*_{\mathcal{X}}}\, \lrcorner\ \mathcal{R}_{\mathcal{X}}(\rho)\,.
\end{align*}
Since the form $\mathcal{R}_{\mathcal{X}}(\rho)$ is canonically related to the identity operator $\mathbb{I}\in \mathrm{End}(D^v(\mathcal{E}_X))$, we obtain
\begin{align*}
{X}_{\nabla(d_1\xi)}|_{\mathcal{E}_{{X}}} \simeq \mathcal{X}_{\nabla(d_1\xi)}|_{\mathcal{E}^*_{\mathcal{X}}}\, \lrcorner\ \mathcal{R}_{\mathcal{X}}(\rho)\,. 
\end{align*}
In other words, the bivector maps the differential of $\mathcal{R}_{{X}}({\xi})$ to the symmetry $-{X}_{\nabla(d_1\xi)}|_{\mathcal{E}_{{X}}}$. Then \textit{the relation between conservation laws and symmetries determined by $\nabla$ is, up to sign, inherited by the bivector corresponding to $\mathcal{R}_{\mathcal{X}}(\mathcal{H}_{\nabla})$}. The reduction of~\eqref{Nablabracket} shows that for $X$-invariant conservation laws $\xi_1, \xi_2 \in E^{0,\hspace{0.1ex} n-1}_1(\mathcal{E})$, the Poisson bracket of their reductions defined by $\mathcal{R}_{\mathcal{X}}(\mathcal{H}_{\nabla})$ represents\footnote{Since we consider the case $n=2$ (and $H^0_{dR}(\mathcal{E}_X) = \mathbb{R}$), the reduction~\eqref{Redofbracketcl} is defined up to an additive real number. In particular, if two $X$-invariant conservation laws of $\mathcal{E}$ commute with respect to $\{\cdot, \cdot\}_{\nabla}$, the inherited Poisson bracket of their reductions is a real number.}
\begin{align}
-\mathcal{R}_{X}(\{\xi_1, \xi_2\}_{\nabla})\,.
\label{Redofbracketcl}
\end{align}
One can say that in this case, the Poisson bracket on $E^{0, \hspace{0.1ex} 0}_1(\mathcal{E}_X)$ defined by $\mathcal{R}_{\mathcal{X}}(\mathcal{H}_{\nabla})$ is \textit{inherited from $\{\cdot, \cdot\}_\nabla$ up to sign}. The Poisson bracket defined by $-\mathcal{R}_{\mathcal{X}}(\mathcal{H}_{\nabla})$ is \textit{inherited from $\{\cdot, \cdot\}_\nabla$}.

\vspace{1ex}

\remarka{The invariant reduction of~\eqref{actionnablagen} shows that the relation between variational $1$-forms and symmetries determined by $\nabla$ is inherited by $\mathcal{R}_{\mathcal{X}}(\mathcal{H}_{\nabla})$.}

\vspace{1ex}

\remarka{Using multi-reduction under a commutative algebra $X$ of symmetries (see~\cite{DrCh2}), one can obtain similar results for $n > 2$ provided the multi-reduction is well-defined.
}

\vspace{1ex}

\remarka{\label{Remindep} The reductions $\mathcal{R}_X(\xi_1)$, $\mathcal{R}_X(\xi_2)$, $\ldots$ of commuting $X$-invariant conservation laws $\xi_1, \xi_2, \ldots \in E^{0, \hspace{0.1ex} n-1}_1(\mathcal{E})$ are independent (i.e., their differentials are linearly independent on a dense open set) if $X_{\nabla(d_1\xi_1)}|_{\mathcal{E}_X}$, $X_{\nabla(d_1\xi_2)}|_{\mathcal{E}_X}$, $\ldots$ are linearly independent on a dense open set.}

\subsection{Reductions of (1+1)-dimensional evolution systems}

Remark~\ref{Remfibers} shows that the existence of the isomorphism between $\mathfrak{T}^*[1]\mathcal{E}_X$ and $\mathcal{E}^*_{\mathcal{X}}$ considered above depends only on some non-degeneracy of the differential form $\mathcal{R}_{\mathcal{X}}(\rho)$. In this section, we prove the existence of such isomorphisms for some reductions of $(1+1)$-dimensional evolution systems and derive an explicit formula for the inherited Poisson bivectors.

Let $\mathcal{E}$ be the infinite prolongation of an evolution system
\begin{align*}
u^1_t = f^1\,,\qquad\ldots\,,\qquad u^m_t = f^m\,.
\end{align*}
Here $\zeta = \pi$. In adapted local coordinates, $F^i = u^i_t - f^i$, while $f^i$ are functions of the independent variables $t$ and $x$, the dependent variables $u^i$, and their $x$-derivatives $u^i_{kx}$ up to some finite order. For $(1+1)$-dimensional evolution systems we also use the notation $u^i_k = u^i_{kx}$.
As intrinsic coordinates on $\mathcal{E}$, we take $t$, $x$, and all variables of the form $u^i_{k}$, including $u^i_0 = u^i$. Similarly, as intrinsic coordinates on $\mathcal{E}^*$, we use $t$, $x$, and all variables of the form $u^i_{k}$, $p_{i\, k} = p_{i\, kx}$.

\vspace{1ex}

\remarka{In fact, here we take $p_{i\, kx}|_{\mathcal{E}^*}$ rather than $p_{i\, kx}$, but the difference between them is negligible for our purposes.
}

\vspace{1ex}

Let $X = E_{\varphi}|_{\mathcal{E}}$ be a symmetry of $\mathcal{E}$. The subsystem $\mathcal{E}_X\subset \mathcal{E}$, defined by the condition $X = 0$, is given by the relations $D_x^{\hspace{0.1ex} k}(\varphi^i)|_{\mathcal{E}} = 0$. Without loss of generality, we assume that the components $\varphi^1, \ldots, \varphi^m$ do not depend on variables of the form $D_{\alpha}(u^i_t)$. Then $E_{\varphi}(F) = l_{\varphi}(F)$ and the lift $\mathcal{X}$ has the characteristic $(\varphi, -l_{\varphi}^{\,*}(p))$. Suppose that for some integers $k_1$, \ldots, $k_m \geqslant -1$, the determinant 
\begin{align*}
\det \left(\dfrac{\partial \varphi^j}{\partial u^i_{k_i+1}}\right)
\end{align*}
never vanishes on the system $\varphi = 0$, while none of the functions $\varphi^j$ depend on derivatives of the form $u^i_{k_i+2}$\,, $u^i_{k_i+3}$\,, $\ldots$
Then, in a neighborhood of any point of either system $\varphi = 0$, $(\varphi, -l_{\varphi}^{\,*}(p)) = (0, 0)$, we can eliminate the highest order $x$-derivatives and use the remaining variables as coordinates on the systems $\mathcal{E}_X$ and $\mathcal{E}^*_{\mathcal{X}}$, respectively.

Let us adopt the notation $\,\overline{\!\theta}^{\hspace{0.1ex} i}_{k} = {\theta}^{\hspace{0.05ex} i}_{kx}|_{\mathcal{E}^*}$. The variational $1$-form $\rho\in E_1^{1,\hspace{0.1ex} 1}(\mathcal{E}^*)$ is represented by
\begin{align*}
\omega_L|_{\mathcal{E}^*} = p_i\hspace{0.1ex} \,\overline{\!\theta}^{\hspace{0.1ex} i}_0\wedge dx + \ldots\wedge dt\,.
\end{align*}
The reduction of $\rho$ can be derived from the condition
\begin{align*}
\mathcal{L}_{\mathcal{X}}(\omega_L|_{\mathcal{E}^*}) = d_0 \vartheta\,.
\end{align*}
Its $dx$-component yields the relation
\begin{align*}
p_j\dfrac{\partial \varphi^j}{\partial u^i_k} \,\overline{\!\theta}^{\hspace{0.1ex} i}_{k} - \sum_k (-1)^k\, \overline{\mathcal{D}}_{x}^{\,k}\Big(p_j\dfrac{\partial \varphi^j}{\partial u^i_k}\Big) \,\overline{\!\theta}^{\hspace{0.1ex} i}_0 = - \mathcal{L}_{\overline{\mathcal{D}}_x}(\vartheta)\,.
\end{align*}
Here $\overline{\mathcal{D}}_x$ denotes the restriction of the total derivative $\mathcal{D}_x$ to $\mathcal{E}^*$. Note that this equation for $\vartheta\in E^{1, \hspace{0.1ex} 0}_0(\mathcal{E}^*)$ has a unique solution.
Integrating by parts, we find that
\begin{align*}
\vartheta
&= -\sum_{k\geqslant 1} \Big(p_j\dfrac{\partial \varphi^j}{\partial u^i_k} \,\overline{\!\theta}^{\hspace{0.1ex} i}_{k-1} - \overline{\mathcal{D}}_x\Big(p_j\dfrac{\partial \varphi^j}{\partial u^i_k}\Big) \,\overline{\!\theta}^{\hspace{0.1ex} i}_{k-2} + \overline{\mathcal{D}}_x^{\hspace{0.2ex} 2}\Big(p_j\dfrac{\partial \varphi^j}{\partial u^i_k}\Big) \,\overline{\!\theta}^{\hspace{0.1ex} i}_{k-3} + \ldots + (-\overline{\mathcal{D}}_x)^{ k-1}\Big(p_j\dfrac{\partial \varphi^j}{\partial u^i_k}\Big) \,\overline{\!\theta}^{\hspace{0.1ex} i}_{0}\Big)\,.
\end{align*}
The coefficients of $\,\overline{\!\theta}^{\hspace{0.1ex} i}_{k_i}$, $\,\overline{\!\theta}^{\hspace{0.1ex} i}_{k_i - 1}$, $\,\overline{\!\theta}^{\hspace{0.1ex} i}_{k_i - 2}$, \ldots, $\,\overline{\!\theta}^{\hspace{0.1ex} i}_{0}$ have the form
\begin{align*}
&-p_j\dfrac{\partial \varphi^j}{\partial u^i_{k_i+1}}\,,\qquad -p_j\dfrac{\partial \varphi^j}{\partial u^i_{k_i}} + \overline{\mathcal{D}}_x\Big(p_j\dfrac{\partial \varphi^j}{\partial u^i_{k_i+1}}\Big)\,,\qquad -p_j\dfrac{\partial \varphi^j}{\partial u^i_{k_i-1}} + \overline{\mathcal{D}}_x\Big(p_j\dfrac{\partial \varphi^j}{\partial u^i_{k_i}}\Big) - \overline{\mathcal{D}}_x^{\hspace{0.2ex} 2}\Big(p_j\dfrac{\partial \varphi^j}{\partial u^i_{k_i+1}}\Big)\,,\\
&\ldots\,,\qquad
-p_j\dfrac{\partial \varphi^j}{\partial u^i_{1}} + \overline{\mathcal{D}}_x\Big(p_j\dfrac{\partial \varphi^j}{\partial u^i_{2}}\Big) - \overline{\mathcal{D}}_x^{\hspace{0.2ex} 2}\Big(p_j\dfrac{\partial \varphi^j}{\partial u^i_{3}}\Big) + \ldots - (-\overline{\mathcal{D}}_x)^{k_i}\Big(p_j\dfrac{\partial \varphi^j}{\partial u^i_{k_i+1}}\Big)\,,
\end{align*}
respectively. Equating the restrictions of these coefficients to the variables $\varpi_{i\, k_i}$, $\varpi_{i\, k_i-1}$, $\varpi_{i\, k_i-2}$, \ldots, $\varpi_{i\, 0}$, respectively,
we obtain the desired isomorphism of $\mathfrak{T}^*[1]\mathcal{E}_X$ and $\mathcal{E}^*_{\mathcal{X}}$. Indeed, the following relations can be solved for the variables $p_j$
\begin{align*}
\varpi_{i\, k_i} = -p_j\dfrac{\partial \varphi^j}{\partial u^i_{k_i+1}}\bigg|_{\mathcal{E}^*_{\mathcal{X}}}
\end{align*}
Here the restriction to $\mathcal{E}^*_{\mathcal{X}}$ can be replaced with the restriction to $\mathcal{E}_X$ or simply to $\varphi = 0$. Then, at the next step, the variables $p_{j\, 1}$ can be determined from the expressions for $\varpi_{i\, k_i - 1}$, and so on. One can see that, in local coordinates under consideration, there exists a unique lift of a vertical derivation from $\mathcal{E}_X$ to $\mathcal{E}_{\mathcal{X}}^*$ preserving $\mathcal{R}_{\mathcal{X}}(\rho)$. Its uniqueness ensures that the lift is well-defined globally. As it follows from the coordinate expression for $\mathcal{R}_{\mathcal{X}}(\rho)$, the operator that contracts such lifts with $\mathcal{R}_{\mathcal{X}}(\rho)$ establishes an isomorphism between the $\mathcal{F}(\mathcal{E}_X)$-modules of vertical derivations on $\mathcal{E}_X$ and degree-$1$ functions on $\mathcal{E}^*_{\mathcal{X}}$. Therefore, it yields the identification of $\mathfrak{T}^*[1]\mathcal{E}_X$ and $\mathcal{E}^*_{\mathcal{X}}$ relating the corresponding $1$-form $\rho_c$ to $\mathcal{R}_{\mathcal{X}}(\rho)$. Its coordinate form is given by the expressions for $\varpi_{i\, k_i}$, $\varpi_{i\, k_i - 1}$, $\ldots,$ $\varpi_{i\, 0}$.

Thus, if $\nabla\colon \widehat{P}(\mathcal{E})\to \varkappa(\mathcal{E})$ is an $X$-invariant Hamiltonian operator, then there is a Poisson bracket on $E^{0, \hspace{0.1ex} 0}_1(\mathcal{E}_X)$, inherited from $\{\cdot, \cdot\}_\nabla$ up to sign. It is defined by $\mathcal{R}_{\mathcal{X}}(\mathcal{H}_{\nabla})$. We obtain

\vspace{1ex}

\propositiona{\label{MainProp} Let $\varphi$ be a characteristic of a symmetry of a $(1+1)$-dimensional evolution system $\mathcal{E}$ such that for some integers $k_1$, \ldots, $k_m \geqslant -1$, the determinant 
\begin{align*}
\det \left(\dfrac{\partial \varphi^j}{\partial u^i_{k_i+1}}\right)
\end{align*}
never vanishes on the system $\varphi = 0$, while none of the functions $\varphi^j$ depend on variables of the form $u^i_t$, $u^i_{k_i+2}$\hspace{0.15ex}, or their total derivatives.
Suppose that $\nabla$ is an $X$-invariant Hamiltonian operator of $\mathcal{E}$, where $X = E_{\varphi}|_{\mathcal{E}}$. Then $\mathcal{R}_{\mathcal{X}}(\mathcal{H}_{\nabla})$ corresponds to a bivector on $\mathcal{E}_X$. It defines a Poisson bracket on $E^{0, \hspace{0.1ex} 0}_1(\mathcal{E}_X)$, which is inherited from $\{\cdot, \cdot\}_\nabla$ up to sign.}

\vspace{1ex}

Continuing with the same local coordinates on $\mathcal{E}_X$ and $\mathcal{E}^*_{\mathcal{X}}$, denote by $w_{j\,0}$ the vector field 
that corresponds to $p_j$, $j = 1, \ldots, m$. Then $p_{j\, 1}$, $p_{j\, 2}$, \ldots correspond to the Lie derivatives $\mathcal{L}_{\widetilde{D}_x} (w_{j\,0})$, $\mathcal{L}_{\widetilde{D}_x}^{\, 2} (w_{j\,0})$, $\ldots$ along $\widetilde{D}_x = D_x|_{\mathcal{E}_X}$. Using this observation, one can derive an explicit formula for the bivector corresponding to $\mathcal{R}_{\mathcal{X}}(\mathcal{H}_{\nabla})$. Namely, the reduction of the formula~\eqref{Hamil} yields
\begin{align*}
\mathcal{R}_{\mathcal{X}}(\mathcal{H}_{\nabla}) = \dfrac{1}{2}s_{\nabla}|_{\mathcal{E}^*_{\mathcal{X}}}\,\lrcorner\, \mathcal{R}_{\mathcal{X}}(\rho)\,.
\end{align*}
Let $\nabla = \nabla^{ij\, k} \,\overline{\!D}_{kx}$, where $\,\overline{\!D}_{kx} = D_{x}^{\hspace{0.1ex} k}|_{\mathcal{E}}$ and $\nabla(\,\overline{\!\psi})^i = \nabla^{ij\, k} \,\overline{\!D}_{kx}(\,\overline{\!\psi}_j)$ for $\overline{\psi} = (\overline{\psi}_1, \ldots, \overline{\psi}_m)$.
Since $\mathcal{R}_{\mathcal{X}}(\rho)$ is canonically related to the identity operator $\mathbb{I}\in \mathrm{End}(D^v(\mathcal{E}_X))$ and
\begin{align*}
s_{\nabla} = \nabla^{ij\, k} p_{j\, k}\, \partial_{u^i_0} + \overline{\mathcal{D}}_x(\nabla^{ij\, k} p_{j\, k}) \partial_{u^i_1} + \ldots\,,
\end{align*}
we find the explicit formula\footnote{In this formula, both indices $r$ are interpreted as upper indices; therefore, summation over them requires an explicit summation sign. Similarly, both indices $i$ are lower indices.}
\begin{align*}
\mathcal{R}_{\mathcal{X}}(\mathcal{H}_{\nabla}) \simeq \dfrac{1}{2} \sum_i \sum_{r\hspace{0.1ex} =\hspace{0.1ex} 0}^{k_i} \mathcal{L}_{\widetilde{D}_x}^{\, r}\big(\hat{\nabla}_{i}^{j\, k} w_{j\, k}\big)\wedge \partial_{u^i_{r}}\,,
\end{align*}
where $w_{j\, k} = \mathcal{L}_{\widetilde{D}_x}^{\, k} (w_{j\,0})$, $\hat{\nabla}_i^{j\, k} = \nabla^{ij\, k}|_{\mathcal{E}_X}$.
Combining this formula and Proposition~\ref{MainProp}, we obtain the following theorem.

\vspace{1ex}

\theorema{\label{MainTheor} Let $\varphi$ be a characteristic of a symmetry $X = E_{\varphi}|_{\mathcal{E}}$ of a $(1+1)$-dimensional evolution system $\mathcal{E}$ with $m$ dependent variables $u^1$, \ldots, $u^m$ such that, for some integers $k_1$, \ldots, $k_m \geqslant -1$, the determinant 
\begin{align*}
\det \left(\dfrac{\partial \varphi^j}{\partial u^i_{k_i+1}}\right)
\end{align*}
never vanishes on the system $\varphi = 0$, and none of the functions $\varphi^j$ depend on variables of the form $u^i_t$, $u^i_{k_i+2}$\hspace{0.15ex}, or their total derivatives.
If $\nabla = \nabla^{ij\, k} \,\overline{\!D}_{kx}$ is an $X$-invariant Hamiltonian operator of $\mathcal{E}$, then on the algebra $E^{0, \hspace{0.1ex} 0}_1(\mathcal{E}_X)$ of constants of $X$-invariant motion, 
there exists a Poisson bracket inherited from $\{\cdot, \cdot\}_\nabla$. 
In coordinates of the form $t$, $x$, $u^i_0$, \ldots, $u^i_{k_i}$ on $\mathcal{E}_X$, the bivector defining this bracket is given by
\begin{align*}
-\dfrac{1}{2} \sum_i \sum_{r\hspace{0.1ex} =\hspace{0.1ex} 0}^{k_i} \mathcal{L}_{\widetilde{D}_x}^{\, r}\big(\hat{\nabla}_{i}^{j\, k} w_{j\, k}\big)\wedge \partial_{u^i_{r}}\,,\qquad \hat{\nabla}_{i}^{j\,k} = \nabla^{ij\, k}|_{\mathcal{E}_X},
\end{align*}
where $\widetilde{D}_x = D_x|_{\mathcal{E}_X}$, $w_{j\, k} = \mathcal{L}_{\widetilde{D}_x}^{\, k} (w_{j\, 0})$, and $w_{j\, 0}$ are defined by the relations
$\partial_{u^i_{k_i}} = -w_{j\, 0}\, \dfrac{\partial \varphi^j}{\partial u^i_{k_i+1}}\bigg|_{\mathcal{E}_{X}}$.
}

\vspace{1ex}

\remarka{If $X$ is a Hamiltonian symmetry corresponding to $\xi\in E^{0, \hspace{0.1ex} 1}_1(\mathcal{E})$, i.e., $X = X_{\nabla(d_1\xi)}$, then $\mathcal{R}_X(\xi)$ is a Casimir function\footnote{The bivector corresponding to $\mathcal{R}_{\mathcal{X}}(\mathcal{H}_{\nabla})$ relates $\mathcal{R}_X(\xi)$ to the symmetry $-X|_{\mathcal{E}_X} = 0$.} for the bivector corresponding to $\mathcal{R}_{\mathcal{X}}(\mathcal{H}_{\nabla})$. As with reductions of other conservation laws, it is nontrivial if and only if the cosymmetry of $\xi$ does not vanish on $\mathcal{E}_X$ (see Remark 7 in~\cite{DrCh1}) provided, for example, that the assumptions of Theorem~\ref{MainTheor} are satisfied.
}

\section{Examples \label{Sectexamp}}

Let us illustrate Theorem~\ref{MainTheor} by the following examples. We also make use of a more suitable index notation below. Some other computations relevant to Section~\ref{SectKdV} are given in Appendix~\ref{App:B}.

\subsection{Kaup--Boussinesq system \label{KBsect}}

Let $\mathcal{E}$ denote the infinite prolongation of the Kaup--Boussinesq system
\begin{align*}
u_t + uu_x + h_x = 0\,,\qquad h_t + u h_x + u_x h + \dfrac{1}{4}u_{xxx} = 0\,.
\end{align*}
Here $u^1 = u$, $u^2 = h$, $F^1 = u_t + uu_x + h_x$, $F^2 = h_t + u h_x + u_x h + \frac{1}{4}u_{xxx}$. Take $t$, $x$, $u$, $h$, $u_x$, $h_x$, $u_{xx}$, $h_{xx}$, $\ldots$ as coordinates on $\mathcal{E}$. The operator
\begin{align*}
\nabla = 
\begin{pmatrix}
0 & \,\overline{\!D}_x\\
\,\overline{\!D}_x & 0
\end{pmatrix}
\end{align*}
is a Hamiltonian operator of the Kaup--Boussinesq (KB) system. The conservation law corresponding to the cosymmetry $\overline{\psi}$ of the KB with the components $(\overline{\psi}_1, \overline{\psi}_2)$,
\begin{align*}
\overline{\psi}_1 = \dfrac{3}{4}u^2h + \dfrac{3}{8}uu_{xx} + \dfrac{3}{16}u_x^2 + \dfrac{3}{4}h^2 + \dfrac{1}{4}h_{xx}\,,\qquad 
\overline{\psi}_2 = \dfrac{3}{2}uh + \dfrac{1}{4}u^3 + \dfrac{1}{4}u_{xx}
\end{align*}
gives rise to the Hamiltonian symmetry $X = E_{\varphi}|_{\mathcal{E}}$, $\varphi = (\varphi^1, \varphi^2)$,
\begin{align*}
\varphi^1 = \dfrac{3}{2}(u h_x + u_x h) + \dfrac{3}{4}u^2 u_x\! + \dfrac{1}{4}u_{xxx},\,  \varphi^2 = \dfrac{3}{4}(2uu_xh + u^2h_x) + \dfrac{3}{8}uu_{xxx} + \dfrac{3}{4}u_xu_{xx} + \dfrac{3}{2}hh_x + \dfrac{1}{4}h_{xxx}.
\end{align*}
Accordingly, the operator $\nabla$ is $X$-invariant. The system $\mathcal{E}_X$ describes $X$-invariant solutions of the Kaup--Boussinesq system. It is given by the infinite prolongation of
\begin{align*}
&u_t + uu_x + h_x = 0\,,\qquad \dfrac{3}{2}(u h_x + u_x h) + \dfrac{3}{4}u^2 u_x + \dfrac{1}{4}u_{xxx} = 0\,,\\
&h_t + u h_x + u_x h + \dfrac{1}{4}u_{xxx} = 0\,,\qquad \dfrac{3}{4}(2uu_xh + u^2h_x) + \dfrac{3}{8}uu_{xxx} + \dfrac{3}{4}u_xu_{xx} + \dfrac{3}{2}hh_x + \dfrac{1}{4}h_{xxx} = 0\,.
\end{align*}
The variables $t$, $x$, $u$, $h$, $u_x$, $h_x$, $u_{xx}$, $h_{xx}$ can be taken as (global) coordinates on $\mathcal{E}_X$.

\subsubsection{The Poisson bivector}

Let us apply Theorem~\ref{MainTheor}. Here $k_1 = k_2 = 2$. The vector fields $w_{1\, 0}$ and $w_{2\, 0}$ satisfy the system
\begin{align*}
\partial_{u_{xx}} = -\dfrac{1}{4} w_{1\, 0} - \dfrac{3}{8}u\, w_{2\, 0}\,,\qquad \partial_{h_{xx}} = -\dfrac{1}{4} w_{2\, 0}\,.
\end{align*}
Accordingly, $w_{1\, 0} = -4\,\partial_{u_{xx}} + 6u\partial_{h_{xx}}$, $w_{2\, 0} = -4\,\partial_{h_{xx}}$. The total derivative $\widetilde{D}_x = D_x|_{\mathcal{E}_X}$ reads
\begin{align*}
\widetilde{D}_x = \partial_x + u_x \partial_u + h_x \partial_h &+ u_{xx}\partial_{u_x} + h_{xx}\partial_{h_x} - (6uh_x + 6u_x h + 3u^2u_x) \partial_{u_{xx}}\\
& + \Big(6u^2h_x + 3uu_xh + \frac{9}{2}u^3u_x - 3u_x u_{xx} - 6hh_x\Big)\partial_{h_{xx}}\,.
\end{align*}
The corresponding fields $w_{1\, 1} = [\widetilde{D}_x, w_{1\, 0}]$, $w_{2\, 1} = [\widetilde{D}_x, w_{2\, 0}]$, $w_{1\, 2} = [\widetilde{D}_x, w_{1\, 1}]$, $\ldots$ have~the~form
\begin{align*}
&w_{1\, 1} = 4\partial_{u_x} - 6u\partial_{h_x} - 6u_x \partial_{h_{xx}}\,,\qquad w_{2\, 1} = 4\partial_{h_x}\,,\\
&w_{1\, 2} = -4\partial_{u} + 6u\partial_h - (24u^2 - 24h)\partial_{u_{xx}} + (18u^3 - 48uh + 6u_{xx})\partial_{h_{xx}}\,,\\
&w_{2\, 2} = -4\partial_{h} + 24u\partial_{u_{xx}} - (24u^2 - 24h)\partial_{h_{xx}}\,,\\
&w_{1\, 3} = 6u_x \partial_h + (24u^2 - 24h)\partial_{u_x} - (18u^3 - 48uh + 6u_{xx})\partial_{h_x} - 36uu_x\partial_{u_{xx}}\,,\\
&w_{2\, 3} = -24u\partial_{u_x} + (24u^2 - 24h)\partial_{h_x} + 36uu_x \partial_{h_{xx}}\,, \qquad \ldots
\end{align*}
Finally, the bivector from Theorem~\ref{MainTheor} reads
\begin{align}
&-\dfrac{1}{2}\big(w_{2\, 1} \wedge \partial_{u} + w_{1\, 1} \wedge \partial_{h} + w_{2\, 2} \wedge \partial_{u_x} + w_{1\, 2} \wedge \partial_{h_x} + w_{2\, 3} \wedge \partial_{u_{xx}} + w_{1\, 3} \wedge \partial_{h_{xx}}\big).
\label{KBbivec}
\end{align}
It defines the Poisson bracket inherited from $\{\cdot, \cdot \}_{\nabla}$.

\subsubsection{Reductions of conservation laws}

Reductions of conservation laws can be obtained through the algorithm provided in~\cite{DrCh1}.
Let us consider the conservation laws of the Kaup--Boussinesq system represented by the differential forms
\begin{align*}
&\omega_1 = uh\hspace{0.2ex} dx - \Big(u^2h + \dfrac{1}{2}h^2 - \dfrac{1}{8}u_x^2 + \dfrac{1}{4}uu_{xx}\Big)dt\,,\\
&\omega_2 = \dfrac{1}{2}\Big(u^2h + h^2 + \dfrac{1}{4}uu_{xx}\Big)dx - \Big(uh^2 + \dfrac{1}{2}u^3h + \dfrac{1}{4}u_{xx}h - \dfrac{1}{8}u_x h_x + \dfrac{1}{4}u^2u_{xx} + \dfrac{1}{8}uh_{xx} \Big)dt\,,\\
&\omega_3 = u\hspace{0.2ex} dx - \Big(h + \dfrac{1}{2}u^2\Big)dt\,,\qquad \omega_4 = h\hspace{0.2ex} dx - \Big(uh + \dfrac{1}{4}u_{xx}\Big)dt\,.
\end{align*}
They are $X$-invariant and Poisson commuting. The conservation laws corresponding to $\omega_3$ and $\omega_4$ are Casimir functions of the Poisson bracket $\{\cdot, \cdot \}_{\nabla}$ on $\mathcal{E}$, i.e., they correspond to the zero symmetry. Then their reductions are Casimir functions for~\eqref{KBbivec}. They have the form
\begin{align*}
I_3 = \dfrac{1}{4}u^3 + \dfrac{3}{2}uh + \dfrac{1}{4}u_{xx}\,,\qquad I_4 = \dfrac{3}{4}u^2 h + \dfrac{3}{8}uu_{xx} + \dfrac{3}{16}u_x^2 + \dfrac{1}{4}h_{xx} + \dfrac{3}{4}h^2
\end{align*}
up to additive constants. The Hamiltonian symmetries corresponding to $\omega_1$ and $\omega_2$ are given by
\begin{align*}
X_1 = \,\overline{\!D}_x - \partial_x\,,\qquad X_2 = \partial_t - \,\overline{\!D}_t\,,
\end{align*}
where $\,\overline{\!D}_t = D_t|_{\mathcal{E}}$. Their restrictions to $\mathcal{E}_X$ are linearly independent on a dense open set. Then the reductions of the conservation laws represented by $\omega_1$ and $\omega_2$ are independent. They are given by
\begin{align*}
I_1 = & \ \dfrac{1}{4}u_{xx}h - \dfrac{1}{4}u_x h_x + \dfrac{3}{4}u^3h + \dfrac{3}{8}u^2u_{xx} + \dfrac{1}{4}uh_{xx} + \dfrac{3}{2}uh^2,\\
I_2 = &- \dfrac{1}{4} uu_x h_x + \dfrac{5}{8} uu_{xx} h - \dfrac{1}{8} h_x^2 + \dfrac{1}{2} h^3 + \dfrac{3}{16} u^3u_{xx} + \dfrac{1}{4} h h_{xx} 
+ \dfrac{3}{8} u^4h + \dfrac{1}{8} u^2h_{xx} \\
&+ \dfrac{1}{4} u_x^2 h + \dfrac{3}{2} u^2 h^2 + \dfrac{1}{32} u_{xx}^2
\end{align*}
up to additive constants. Since the Poisson bracket defined by the bivector~\eqref{KBbivec} is inherited from $\{\cdot, \cdot \}_{\nabla}$, these constants of $X$-invariant motion correspond to the symmetries
\begin{align*}
X_1|_{\mathcal{E}_X} = \widetilde{D}_x - \partial_x\,, \qquad X_2|_{\mathcal{E}_X} = \partial_t - \widetilde{D}_t
\end{align*}
of $\mathcal{E}_X$, where $\widetilde{D}_t = \,\overline{\!D}_t|_{\mathcal{E}_X}$. Let us recall that the restrictions of $I_1$, $I_2$, $I_3$, and $I_4$ to $X$-invariant KB solutions with connected domains are real numbers.

\vspace{1ex}

\remarka{The bivector~\eqref{KBbivec}, together with the independent Poisson commuting constants of $X$-invariant motion $I_1$, $I_2$, $I_3$, and $I_4$, makes the system $\mathcal{E}_X$ integrable.}

\subsection{KdV \label{SectKdV}}
Let us demonstrate the reduction of a higher-order Hamiltonian operator.

Denote by $\mathcal{E}$ the infinite prolongation of the Korteweg--de Vries equation
\begin{align*}
u_t = 6uu_x + u_{xxx}\,.
\end{align*}
Here $u^1 = u$, $F = u_t - 6uu_x - u_{xxx}$. We take $t$, $x$, $u$, $u_x$, $u_{xx}$, $\ldots$ as coordinates on $\mathcal{E}$.
The operator
\begin{align*}
\nabla = \,\overline{\!D}_x^{\,3} + 4u \,\overline{\!D}_x + 2u_x
\end{align*}
is a Hamiltonian operator of the KdV, where $\,\overline{\!D}_x = D_x|_{\mathcal{E}}$. The conservation law corresponding to the cosymmetry $\overline{\psi}$ with the component $u_{xx} + 3u^2$ gives rise to the Hamiltonian symmetry $X = E_{\varphi}|_{\mathcal{E}}$,
\begin{align*}
\varphi = u_{5x} + 10uu_{xxx} + 20u_xu_{xx} + 30u^2u_x\,.
\end{align*}
Here $u_{5x} = u_{xxxxx}$. Accordingly, the operator $\nabla$ is $X$-invariant. The corresponding system $\mathcal{E}_X$ describes $X$-invariant solutions of the KdV. It is given by the infinite prolongation of
\begin{align*}
u_t = 6uu_x + u_{xxx}\,,\qquad u_{5x} + 10uu_{xxx} + 20u_xu_{xx} + 30u^2u_x = 0\,.
\end{align*}
The variables $t$, $x$, $u$, $u_x$, $u_{xx}$, $u_{xxx}$, $u_{4x} = u_{xxxx}$ can be taken as (global) coordinates on $\mathcal{E}_X$.

\subsubsection{The Poisson bivector}

Let us apply Theorem~\ref{MainTheor}. Here $\varphi^1 = \varphi$, $k_1 = 4$,
\begin{align*}
\dfrac{\partial \varphi^1}{\partial u^1_{k_1 + 1}} = 1\,, \ \ \widetilde{D}_x = \partial_x + u_x\partial_u + u_{xx}\partial_{u_x} + u_{xxx}\partial_{u_{xx}} + u_{4x}\partial_{u_{xxx}} - (10uu_{xxx} + 20u_xu_{xx} + 30u^2u_x)\partial_{u_{4x}}.
\end{align*}
The corresponding vector fields $w_{1\, 0}$, $w_{1\, 1} = [\widetilde{D}_x, w_{1\, 0}]$, $w_{1\, 2} = [\widetilde{D}_x, w_{1\, 1}]$, $\ldots$ have the form
\begin{align*}
&w_{1\, 0} = -\partial_{u_{4x}},\ \ w_{1\, 1} = \partial_{u_{xxx}}\,,\ \ w_{1\, 2} = -\partial_{u_{xx}} + 10u\partial_{u_{4x}},\ \
w_{1\, 3} = \partial_{u_x} - 10u\partial_{u_{xxx}} - 10u_x \partial_{u_{4x}},\\
&w_{1\, 4} = -\partial_u + 10u \partial_{u_{xx}} + (10u_{xx} - 70u^2) \partial_{u_{4x}}, \, w_{1\, 5} = -10u \partial_{u_x} + 10u_x \partial_{u_{xx}} - (10u_{xx} - 70u^2) \partial_{u_{xxx}}, \, \ldots
\end{align*}
The bivector from Theorem~\ref{MainTheor} reads
\begin{align*}
&-\dfrac{1}{2}\big((w_{1\, 3} + 4uw_{1\, 1} + 2u_x w_{1\, 0})\wedge \partial_{u} + \ldots + \mathcal{L}_{\widetilde{D}_x}^{\, 4} (w_{1\, 3} + 4uw_{1\, 1} + 2u_x w_{1\, 0})\wedge \partial_{u_{4x}} \big)\\
\notag
&= \partial_u\wedge (\partial_{u_x} - 6u\partial_{u_{xxx}} - 12u_x \partial_{u_{4x}}) + \partial_{u_x}\wedge (6u\partial_{u_{xx}} + 6u_x\partial_{u_{xxx}} + (8u_{xx} - 30u^2)\partial_{u_{4x}})\\
\notag
&\ \ - \partial_{u_{xx}}\wedge ((2u_{xx}\! -\! 30u^2)\partial_{u_{xxx}}\! + (2u_{xxx}\! - 60uu_x)\partial_{u_{4x}}) +
\partial_{u_{xxx}}\wedge (120u^3\! - \! 80uu_{xx}\! + 60u_x^2\! - 2u_{4x})\partial_{u_{4x}}.
\end{align*}
It defines the Poisson bracket inherited from $\{\cdot, \cdot \}_{\nabla}$. A complete set of Poisson commuting, independent constants of $X$-invariant motion, arising as reductions of conservation laws, is presented in~\cite{DrCh1}.

\vspace{1ex}

\remarka{The KdV admits the Hamiltonian operator $D_x|_{\mathcal{E}}$, which is also $X$-invariant. A different approach to reduction of  operators of this form was considered, for example, in~\cite{BogoNov, NoMaPiZa}.
}

\vspace{1ex}

\remarka{
The operators $D_x|_{\mathcal{E}}$ and $\nabla$ of the KdV form a Poisson pencil. Hence, their invariant reductions also form a Poisson pencil. The invariant reduction of $D_x|_{\mathcal{E}}$ results in the bivector}
$$
\dfrac{1}{2}\big(w_{1\, 1}\wedge\partial_u + w_{1\, 2}\wedge\partial_{u_x} + w_{1\, 3}\wedge\partial_{u_{xx}} + w_{1\, 4}\wedge\partial_{u_{xxx}} + w_{1\, 5}\wedge\partial_{u_{4x}}\big)\,.
$$

\section{Conclusion}

The mechanism of invariant reduction provides a natural framework for understanding how symmetry reductions inherit invariant geometric structures. In this paper, we demonstrated that this mechanism also applies to local Hamiltonian operators: under suitable conditions, their reductions give rise to Poisson brackets on the reduced systems.
While our focus was on finite-dimensional symmetry reductions, the mechanism is not limited to the finite-dimensional setting. Interpretation of reductions of Hamiltonian operators to infinite-dimensional systems can be based on a direct analog of the formula~\eqref{actionnablagenomeg}. Reductions of the canonical presymplectic structures of cotangent equations can be described in terms of total differential operators, using the approach via compatibility complexes~\cite{Ver} suitably adapted to the graded-commutative case.
Further investigation of these structures and their properties warrants separate study, particularly in the context of multidimensional Hamiltonian systems, including Lagrangian ones.

It is worth mentioning that finite-dimensional reductions of integrable Hamiltonian systems may inherit integrability via the invariant reduction mechanism. This is illustrated in Section~\ref{Sectexamp} by reductions of the Kaup--Boussinesq system and the KdV equation.
The dynamics of such~reductions is governed by their Cartan distributions and, in general, may not be captured by families of compatible vector fields on the fibers of the underlying submersions. It would be of interest to find, for instance, physical examples of integrable reductions whose underlying submersions are nontrivial bundles. Another potentially interesting direction is to study the interaction of the inherited Poisson bivectors with invariant solutions, including the characterization of solutions on which their ranks drop.

Finally, let us note a related yet distinct important problem: defining cotangent equations for arbitrary sufficiently regular PDEs in terms of their intrinsic geometry. This problem appears to be highly nontrivial, but its solution would provide further insight into the geometric origin of Poisson brackets arising in the theory of PDEs.

\vspace{3ex}

\centerline{\bf{\Large Acknowledgments}}

\vspace{2ex}

The author gratefully acknowledges Alexander~Verbovetsky for suggesting the idea of reducing Poisson brackets by interpreting them as conservation laws of the cotangent equations.
He also thanks Eric Boulter and Raphaël Belliard for valuable discussions, the University of Saskatchewan for hospitality, Prof. Alexey Shevyakov for financial support through the NSERC grant RGPIN 04308-2024, and the Pacific Institute for the Mathematical Sciences for support through a PIMS Postdoctoral Fellowship.

\section*{Data availability}
No data was used for the research described in the article.

\appendix

\section{Elements of the graded-commutative geometry of $J^{\infty}(\hat{\pi})$}\label{App:A}

The algebra $\mathcal{F}(\hat{\pi})$ is non-negatively $\mathbb{Z}$-graded. It can be identified with the exterior algebra of the $\mathcal{F}(\pi)$-module of $\mathcal{C}$-differential operators $\widehat{P}(\pi)\to \mathcal{F}(\pi)$. This identification gives rise to the inclusion $\mathcal{F}(\pi)\subset \mathcal{F}(\hat{\pi})$, which, in
geometric terms, corresponds to the pullback along the projection~\eqref{proj}.

All differential forms on $J^{\infty}(\hat{\pi})$ are polynomial in the variables $p_{i\,\alpha}$. The algebra $\Lambda^{*}(\hat{\pi})$ is bigraded, with the bigrading assigned as follows:
\begin{align*}
x^i(0, 0)\,,\qquad u^i_{\alpha}(0, 0)\,,\qquad p_{i\,\alpha}(1, 0)\,,\qquad dx^i(0, 1)\,,\qquad du^i_{\alpha}(0, 1)\,,\qquad dp_{i\,\alpha}(1, 1)\,.
\end{align*}
The first component is the internal degree (inherited from $\mathcal{F}(\hat{\pi})$), and the second is the differential form degree. Similarly, the projection~\eqref{proj} induces the inclusion $\Lambda^*(\pi)\subset \Lambda^*(\hat{\pi})$.

We denote the internal degree by $|\cdot|$; for instance, $|u^i_{\alpha}| = 0$, $|p_{i\, \alpha}| = 1$, $|du^{i}_{\alpha}| = 0$, $|dp_{i\, \alpha}| = 1$. The signs in algebraic expressions are governed by the inner product of the bigradings. In particular,
\begin{align*}
p_i\hspace{0.15ex} du^j = du^j\hspace{0.15ex} p_i\,,\quad p_i\hspace{0.15ex} dp_j = - dp_j\hspace{0.2ex} p_i\,,\quad du^i\wedge dp_j = - dp_j\wedge du^i\,,\quad dp_{i}\wedge dp_j = dp_j\wedge dp_i\,.
\end{align*}

For vector fields $X$, $Y$ (graded derivations of $\mathcal{F}(\hat{\pi})$) and any differential form $\omega\in \Lambda^*(\hat{\pi})$, one~has
\begin{align*}
\mathcal{L}_{X}\, \omega = d(X \lrcorner\, \omega) + X \lrcorner\, d\omega \,,\qquad \mathcal{L}_{X}(Y\lrcorner\, \omega) = [X, Y]\, \lrcorner\, \omega + (-1)^{|X|\cdot |Y|} Y\lrcorner\, (\mathcal{L}_{X} \omega)\,.
\end{align*}
The contraction of vector fields with differential forms is defined by bringing partial derivatives and differentials together, with vector fields placed on the left. For example, if $\omega$ is a differential $1$-form of the form $du^i_{\alpha}\, \omega^{\alpha}_i + dp_{i\, \alpha}\, \omega^{i\, \alpha}$ and $X = X^i_{\alpha} \partial_{u^i_{\alpha}} + X_{i\, \alpha} \partial_{p_{i\, \alpha}}$, then
\begin{align*}
X\lrcorner\, \omega = X^i_{\alpha}\omega^{\alpha}_i + X_{i\, \alpha}\omega^{i\, \alpha}\,.
\end{align*}

The ideal $I_{\mathcal{E}^*}$ of the system $\mathcal{E}^*\subset J^{\infty}(\hat{\pi})$ consists of functions of the form $\square(l_F^{\hspace{0.15ex} *}(p), F)$, where  $\square\colon \widehat{\varkappa}(\hat{\pi})\to \mathcal{F}(\hat{\pi})$ is a $\mathcal{C}$-differential operator.

\section{Some other computations relevant to Section~\ref{SectKdV}}\label{App:B}

The linearization of $F$ has the form $l_F = D_t - 6uD_x - 6u_x - D_{x}^{\hspace{0.15ex} 3}$. Then $l_F^{\hspace{0.15ex} *} = -D_t + 6uD_x + D_{x}^{\hspace{0.15ex} 3}$ and the cotangent system $\mathcal{E}^*$ is given by the infinite prolongation of
\begin{align*}
&-p_t + 6up_x + p_{xxx} = 0\,,\qquad u_t - 6uu_x - u_{xxx} = 0\,.
\end{align*}
We regard $t$, $x$, $u$, $p$, $u_x$, $p_x$, $u_{xx}$, $p_{xx}$, $\ldots$ as coordinates on $\mathcal{E}^*$. One can take $\nabla_{e} = D_x^{\hspace{0.15ex} 3} + 4u D_x + 2u_x$ as an extension of the $\nabla$ to jets. The formula~\eqref{bivec} takes the form
\begin{align*}
l_F\circ \nabla_{e} - \nabla_{e}^{\hspace{0.15ex} *}\circ l_F^{\hspace{0.15ex} *} = 4FD_x + 2D_x(F)\,.
\end{align*}
Therefore, $\Delta_{\psi}(F) = 4FD_x(\psi) + 2\psi D_x(F)$. Let us put $\Delta_{\psi} = 4D_x(\psi) + 2\psi D_x$. Then $\Delta_{p}^{*}(p) = 4p_x p$. The corresponding degree-$1$ symmetry of $\mathcal{E}^*$ reads
\begin{align*}
&s_{\nabla} = (p_{xxx} + 4up_x + 2u_x p) \partial_u + 2pp_x\partial_p + (p_{4x} + 6u_xp_x + 4up_{xx} + 2u_{xx} p) \partial_{u_x} + 2pp_{xx}\partial_{p_x} + \ldots
\end{align*}
Here $p_{4x} = p_{xxxx}$. As $\omega_L$ in~\eqref{Green}, one can take
\begin{align*}
\omega_L = p\, \theta \wedge dx + \big(6up\, \theta + p\, \theta_{xx} - p_x \theta_x + p_{xx} \theta \big)\wedge dt\,,
\end{align*}
where $\theta = \theta^1 = du - u_x dx - u_t\hspace{0.15ex} dt$, $\theta_x = \theta^1_x = du_x - u_{xx} dx - u_{tx} dt$, $\ldots$
The conservation law $\mathcal{H}_{\nabla}$ from~\eqref{Hamil} is represented by
\begin{align*}
-\dfrac{1}{2} s_{\nabla}\lrcorner\, &(\omega_L|_{\mathcal{E}^*}) =  \dfrac{1}{2}\, p (p_{xxx} + 4up_x) dx\\
&+\dfrac{1}{2}\big(10up p_{xxx} + 24u^2pp_x + p(p_{5x} + 10u_{xx}p_x + 8u_xp_{xx}) - p_x (p_{4x} + 8up_{xx}) +
p_{xx} p_{xxx} \big) dt\,.
\end{align*}

The lift of $X$ to the cotangent equation is the following degree-$0$ symmetry
\begin{align*}
\mathcal{X} = \varphi \hspace{0.15ex} \partial_u - l^{\hspace{0.15ex} *}_{\varphi}(p)\hspace{0.15ex} \partial_p + \ldots\,,\qquad l^{\hspace{0.15ex} *}_{\varphi}(p) = -p_{5x} - 10up_{xxx} - 10u_xp_{xx} - 10u_{xx}p_x - 30u^2 p_x\,.
\end{align*} 
Since $X$ is a Hamiltonian symmetry for the operator $\nabla$, the conservation law $\mathcal{H}_{\nabla}$ is $\mathcal{X}$-invariant and $[s_{\nabla}, \mathcal{X}] = 0$.
The system $\mathcal{E}^*_{\mathcal{X}}$ is the infinite prolongation of
\begin{align*}
&u_t = 6uu_x + u_{xxx}\,, &&u_{5x} + 10uu_{xxx} + 20u_xu_{xx} + 30u^2u_x = 0\,,\\
&p_t = 6up_x + p_{xxx}\,, &&p_{5x} + 10up_{xxx} + 10u_xp_{xx} + 10u_{xx}p_x + 30u^2 p_x = 0\,.
\end{align*}

The reduction of the canonical variational $1$-form $\rho$ is given by the differential form $\vartheta|_{\mathcal{E}^*_{\mathcal{X}}}$, where
\begin{align*}
\mathcal{L}_{\mathcal{X}}(\omega_L|_{\mathcal{E}^*}) = d_0 \vartheta\,.
\end{align*}
From the $dx$-component of this equation, one unambiguously finds
\begin{align*}
\vartheta = \, & -p\ \overline{\!\theta}_{4x} + p_{x}\,\overline{\!\theta}_{xxx} - (10up + p_{xx})\,\overline{\!\theta}_{xx} + (-10u_x p + 10up_x + p_{xxx})\,\overline{\!\theta}_{x} {}\\
&- (10up_{xx} + (10u_{xx} + 30u^2)p + p_{4x})\,\overline{\!\theta}\,,
\end{align*}
where $\,\overline{\!\theta} = \theta|_{\mathcal{E}^*}$, $\,\overline{\!\theta}_x = \theta_x|_{\mathcal{E}^*}$, $\ldots$

\end{document}